\documentstyle[aps,epsfig]{revtex}
\begin{document}
\title{Simultaneous near-field and far field spatial quantum correlations\\
        in spontaneous parametric down-conversion}
\author{E.~Brambilla, A.~Gatti, M.~Bache and L.~A.~Lugiato}
\address{INFM,Dipartimento di Scienze CC.FF.MM.,
Universit\`a dell'Insubria, Via Valleggio 11, 22100 Como, Italy}
\maketitle
\begin{abstract}
We study the spatial correlations of quantum fluctuations that can be
observed in multi-mode spontaneous parametric down-conversion in the regime of high gain.
A stochastic model has been solved numerically to obtain quantitative results
beyond the stationary plane-wave pump approximation.
The pulsed shape of the pump beam and other features of the system, such as
spatial walk-off and diffraction are taken into account.
Their effect on the spatial quantum correlations predicted by the plane-wave pump
theory is investigated, both for near field and far field measurements, in a type I
and in a type II phase-matching configuration.
\end{abstract}
\centerline{Version \today}

\section{Introduction}
The spatial aspects of quantum optical fluctuations have been the
object of several studies in the past \cite{advances,review,aspects}.
In general they show up in nonlinear optical processes, typically wave-mixing phenomena
which involve a large number of spatial modes of the electromagnetic field.
Recently there has been a renewal of attention because
of new potential applications which exploit the quantum properties of the field
for image processing or multi-channel operations.
Examples are quantum holography \cite{holo}, the quantum teleportation of optical images \cite{telep}, and
the measurement of small displacements beyond the Rayleigh limit\cite{fabre}.
An overview of this relatively new branch of quantum
optics, for which the name {\it quantum imaging} was coined, can be found in \cite{qimages}.

The process of frequency down-conversion is particularly suitable for this kind
of applications because of its large emission bandwidth in the spatial
frequency domain. We consider spontaneous parametric down-conversion (SPDC) taking place in a crystal with a
second-order nonlinearity set in a travelling-wave configuration. In
this process the photons of a high intensity pump field are split
into pairs of photons of lower energy and momentum through the
nonlinear interaction with the medium. Since no signal field is
injected, down-conversion is initiated only by vacuum fluctuations
that equally cover all spatial and temporal frequencies. The
spontaneous fluorescence pattern that arises has therefore the
angular spectrum determined by phase-matching conditions, and
depends only on the linear dispersion properties of
the nonlinear material.

Recently, we used multi-mode theory in order to demonstrate that SPDC
is able to display spatial quantum correlation
effects in the far field zone, where the particle-like character
of  the generated field has its clearest manifestation
\cite{prl,fluor}. In particular, assuming that the pump field is a
stationary plane-wave, the theory predicts noise
reduction well below the shot noise level for the
difference in the number of photons measured from two detection
areas $R_1$ and $R_2$ corresponding to couples of phase-conjugate
(signal and idler) modes. In other terms, the photon number
measured over the two detection areas are identical
even at the quantum level. This phenomenon finds its explanation in
the conservation of the photon transverse momentum which is fulfilled in each
elementary down-conversion process: for each photon detected in
say area $R_1$, the detection of its twin in $R_2$ is ensured by
this law.

It is important to stress that
these result holds both in a low and in a high gain regime. However, only in the first case
single photon pairs can be resolved in time by the detectors
and information on spatial quantum correlations can be obtained
directly from coincidence measurements.
In this paper we shall refer more explicitly to the second case,
where a large number of photons are emitted in each mode
and their detection gives rise to continuous photocurrents.
The observation of photon number correlation phenomena in such a regime is
the aim of an experiment presently performed at the University of
Insubria at Como. In this experiment photodetection is performed by means of
a high quantum efficiency charged coupled device (CCD) camera, which is able to resolve
photon number fluctuations that are below the standard quantum limit \cite{CCD}.
The pump field is a high power picosecond laser pulse that provides
energy for a large number of down-converted photons, in a configuration
such that the plane-wave
and continuous-wave (cw) pump approximations are very raw.

In this paper we present
a realistic description of the  system, based on
a numerical model that includes the finite frequency
bandwidth of the pump,
both in spatial and temporal frequency domains.
Other features of the system that are relevant
from an experimental point of view, such as spatial and temporal walk-off, different kinds of
linear dispersion and phase-matching (type I and type II
crystals), are included in the model. It is important to
investigate how they affect the spatial quantum
correlation phenomena predicted by the plane-wave pump theory,
also in order to identify the best conditions under which they
can be observed in the experiment.
A numerical evaluation of the far field photon number correlation function
is presented in \cite{lantznum} in the case of a type I crystal at degeneracy.
In \cite{lantznum}, which treat spontaneous down-conversion
within a classical framework, the shape of the pump pulse included into the numerical model
is taken from experimental data and the obtained signal-idler correlation peak displayed
between symmetrical point reproduce well the correlation measured experimentally.

We shall also focus our attention on the spatial correlation property of
SPDC in the near field, where the signal and idler beams are found to
exhibit quantum correlated photon number fluctuations when measured from detection areas that image the same portion
of the beam cross-section.
Twin photons are indeed generated simultaneously and
they remain localized in a limited region
of space as a long as they are observed close to the crystal.
This "position entanglement" of the generated photon pairs
can be seen as the near field counterpart of the momentum entanglement
which can be observed in the far field.
However, we shall see that for a realistic crystal length the measurement of near field correlation
is strongly affected by propagation effects, in particular diffraction and spatial walk-off.
We shall propose a procedure to overcome at least partially this problem.

The paper is organized as follows.
In Sec.\ref{model} we briefly introduce the theoretical model
used to describe SPDC within a classical framework.
The quantum description of the system is illustrated in Sec.\ref{quantdescr},
where a fully analytical treatment is developed in the
framework of the plane-wave and cw pump approximation (PWPA).
It is based on a multi-mode input-output formalism, first introduced in \cite{kolobov89}
for a type I crystals at degeneracy, which
is here extended to a type II phase-matching configuration.

In Sec.\ref{meanint} we give a qualitative description of the phase-matching
mechanism that determines both the photon number distribution
and the characteristic bandwidths of the down-converted field, illustrating
thereby the differences between type I and type II phase-matching.

In Sec.\ref{correlations} we define the quantities that can be
measured experimentally and that put in evidence
the quantum nature of the spatial correlations in which we are interested.
Their analytical expressions are derived within the PWPA,
which will be used to interpret the results of the numerical model.

The last part of the paper (Sec.\ref{nummodel})
is devoted to present the numerical results obtained for two
particular crystals with different phase-matching (type I and type II).
The amount of correlations that may be achieved is evaluated as a function of different
parameters that can be varied experimentally, such as the size of
the pump beam waist and the size of the detectors.

\section{Classical description of the process}\label{model}
We decompose the electric field in the superposition of three
quasi-monochromatic wavepackets (denoted with $E_0$, $E_1$ and $E_2$)
of central frequencies $\omega_0$, $\omega_1$ and $\omega_2$,
corresponding to the pump, the signal and the idler
fields, respectively. These frequencies are taken to satisfy
the energy conservation condition $\omega_1+\omega_2=\omega_0$.
Assuming the mean direction of propagation is the $z$ direction, and denoting
with $\vec{x}=(x,y)$ the coordinate vector in the transverse plane, we can write
\begin{eqnarray}
\label{mwaves}
E_j(z,\vec{x},t)\propto A_j(z,\vec{x},t)~e^{ik_jz-i\omega_jt}+c.c.
\hspace{1 cm}
(j=0,1,2)\;.
\end{eqnarray}
where $k_j=n_j\omega_j/c$ is the wave number
of wave $j$ at the carrier frequency along the $z$-axis (for an
extraordinary wave the refraction index $n_j$ depends on the propagation direction,
a property leading to spatial walk-off).
To simplify the notation we have ignored the vectorial
character of the three fields,
their polarization being determined by the kind of phase-matching condition
that are met inside the crystal.

Within the paraxial and slowly varying envelope approximation,
the propagation equations for the signal and idler (S/I) field envelopes and the pump field
envelope can be written in the form \cite{handbook}
\begin{mathletters}
\label{qwaveq}
\begin{eqnarray}
& &\frac{\partial A_j}{\partial z}+k_j'\frac{\partial A_j}{\partial t}
+\frac{i}{2}k_j''\frac{\partial^2 A_j}{\partial t^2}
-\rho_j \frac{\partial A_j}{\partial y}
-\frac{i}{2k_j}\nabla_{\perp}^2 A_j
=\sigma A_0 A_l^*e^{-i\Delta_0z}
\hspace{1cm}(j,l=1,2;\;j\neq l)\;,\label{dAja}\\
& &\frac{\partial A_0}{\partial z}+k_0'\frac{\partial A_0}{\partial t}
+\frac{i}{2}k_0''\frac{\partial^2 A_0}{\partial t^2}
-\rho_0 \frac{\partial A_0}{\partial y}
-\frac{i}{2k_0}\nabla_{\perp}^2 A_0
=-\sigma A_1 A_2 e^{i\Delta_0z}\label{dAjb}
\end{eqnarray}
\end{mathletters}
The driving terms on the r.h.s. describe the wave-mixing process
due to the second-order nonlinearity of the medium, the coupling
constant $\sigma$ being proportional to the effective second-order
susceptibility $\chi^{(2)}_{eff}$ characterizing the down-conversion
process. $\Delta_0=k_1+k_2-k_0$ is the collinear phase-mismatch of
the central frequency components.

Linear propagation is described by the l.h.s. of these equations:
the terms proportional to
$k_j'=\left(\frac{\partial k_j}{\partial\omega}\right)_{\omega=\omega_j}$ and
$k_j''=\left(\frac{\partial^2 k_j}{\partial\omega^2}\right)_{\omega=\omega_j}$ lead to
temporal walk-off between the different waves and group velocity dispersion respectively,
while the terms containing the first and second order derivatives
in the transverse coordinates $(x,y)$
are responsible of spatial walk-off and diffraction respectively.
$\rho_j$
indicates the walk-off angle of wave $j$, determined by the
anisotropy of the crystal (the walk-off direction is taken
along the $y$-axis). Linear losses are neglected,
so that the three wave exchange energy but their total energy is conserved.

In Sec.\ref{numtypeI}
we shall also consider the special case of a type I phase-matched crystal
where the signal and the idler fields have the same polarization
and are observed close to the
degenerate frequency $\omega_1=\omega_2=\omega_0/2$.
Under these conditions the signal and idler fields are no more distinguishable
and the down-converted field must be described by a single slowly-varying envelope
$A(z,\vec{x},t)$ satisfying the following propagation equation
\begin{eqnarray}
\label{dA}
\frac{\partial A}{\partial z}+k'\frac{\partial A}{\partial t}
+\frac{i}{2}k''\frac{\partial^2 A}{\partial t^2}
-\rho_1 \frac{\partial A}{\partial y}
-\frac{i}{2k}\nabla_{\perp}^2 A
=\sigma A_0 A^*e^{-i\Delta_0z}\;,
\end{eqnarray}
which is readily obtained from Eqs.(\ref{dAja}) by dropping
the S/I indexes $j,l$, which denote different polarizations and/or carrier frequencies
in the non-degenerate case.

In a single-pass configuration with crystal length on the
order of a few millimeters, the amplitudes of the down-converted
field remain small with respect to the pump amplitude and the
nonlinear driving term in the r.h.s of Eq.(\ref{dAjb}) can
be neglected. The pump depletion due to down-conversion and
absorption is indeed of small entity, unless extremely high
intensity laser sources are used. We shall therefore work within
the parametric approximation, that treats the pump as a
known classical field which propagates linearly inside the
crystal, while the down-converted fields are quantized according
to the rule that are briefly illustrated in the next section.

\section{Quantum description in the parametric approximation}\label{quantdescr}
We need now to substitute the classical signal and idler fields
with operators. Making the formal substitution for the field
envelopes $A_j(z,\vec{x},t) \rightarrow
a_j(z,\vec{x},t)\;,(j=1,2),$ we impose the following commutation
rules at equal $z$ \cite{review}
\begin{eqnarray}
\label{commut}
\left[a_i(z,\vec{x},t),a_j^{\dag}(z,\vec{x}~',t')\right]
&=&\delta_{ij}\delta(\vec{x}-\vec{x}~')\delta(t-t')\;,\\
\left[a_i(z,\vec{x},t),a_j(z,\vec{x}~',t')\right]&=&0~~~~~(i,j=1,2)\nonumber\;,
\end{eqnarray}
valid within the framework of the paraxial and quasi-monochromatic
approximations. With this definition
\begin{equation}
\label{flux}
I_j(z,\vec{x},t)=a_j^{\dag}(z,\vec{x},t)a_j(z,\vec{x},t)~~~~~(j=1,2),
\end{equation}
is the photon flux density operator associated to wave $j$: its expectation value
gives the mean number of photons crossing a region of unit area in
the transverse plane.
In the linear regime the field operators obey the same equations
as the corresponding classical quantities.
To our purposes, it is useful to introduce the Fourier transforms of the
field envelopes with respect
to time and to the transverse plane coordinates:
\begin{equation}
a_j(z,\vec{q},\Omega)=
     \int\frac{d\vec{x}}{2\pi}\int\frac{dt}{\sqrt{2\pi}}
                 a_j(z,\vec{x},t)e^{-i\vec{q}\cdot\vec{x}+i\Omega t}
     \hspace{1cm}(j=1,2)\;.
\end{equation}
A similar definition holds also for the Fourier component
$A_0(z,\vec{q},\Omega)$ of the classical pump field envelope. The
propagation equations (\ref{dAja}) take then the form
\begin{eqnarray}
\label{waveq}
\frac{\partial a_j(z,\vec{q},\Omega)}{\partial z}&=&
    i\delta_j(\vec{q},\Omega)
    a_j(z,\vec{q},\Omega)\\
    &+&\sigma e^{-i\Delta_0 z}
    \int\frac{d\vec{q}~'}{2\pi} \int\frac{d\Omega'}{\sqrt{2\pi}}
         A_0(z,\vec{q}-\vec{q}~',\Omega-\Omega')a_l^{\dag}(z,-\vec{q}~',-\Omega')
    \hspace{1cm}(j,l=1,2;\;j\neq l)\nonumber\;,
\end{eqnarray}
where
\begin{equation}
\label{kz}
\delta_j(\vec{q},\Omega)
=k_j'\Omega+\frac{1}{2}k_j''\Omega^2+\rho_j q_y-\frac{1}{2k_j}(q_x^2+q_y^2)\;,~~~~~(j=1,2),
\end{equation}
is the quadratic expansion of
$k_{jz}(\omega_j+\Omega,\vec{q})-k_j$ around $\vec{q}=0,\Omega=0$, and
$k_{jz}(\omega_j+\Omega,\vec{q})=\sqrt{k_j^2(\omega_j+\Omega,\vec{q})-q^2}$
denotes the $z$-component of the $k$-vector associated to the
$(\vec{q},\Omega)_j$ plane-wave mode. In particular the walk-off
angle $\rho_j$ can be identified as
$\frac{\partial k_j}{\partial q_y}$ calculated for $\vec{q}=0,\Omega=0$.

Eqs.(\ref{waveq}) contain the convolution integral in Fourier space
of the S/I field envelope with the pump field envelope.
Within the undepleted pump approximation, the latter can be expressed as
\begin{mathletters}
\label{prop0}
\begin{eqnarray}
&&A_0(z,\vec{q},\Omega)=e^{i\delta_0(\vec{q},\Omega)z}A_0(z=0,\vec{q},\Omega)\;,\\
&&\delta_0(\vec{q},\Omega)
=k_0'\Omega+\frac{1}{2}k_0''\Omega^2+\rho_0 q_y-\frac{1}{2k_0}(q_x^2+q_y^2)\;,~~~~~(j=1,2),
\end{eqnarray}
\end{mathletters}
the $z=0$ plane being taken at the input face of the crystal.
In the following we shall assume that the pump pulse has a Gaussian
profile both in space and time, of beam waist $w_0$ and
time duration $\tau_0$ at $z=0$:
\begin{equation}
A_0(z=0,\vec{x},t)=(2\pi)^{3/2}A_p e^{-(x^2+y^2)/w_0^2}e^{-t^2/\tau_0^2}\;.
\end{equation}
In Fourier space we have then the expression
\begin{equation}
\label{pump0}
A_0(z=0,\vec{q},\Omega)=2\sqrt{2}\frac{A_p}{\delta q_0^2\delta \omega_0}
e^{-(q_x^2+q_y^2)/\delta q_0^2}e^{-\Omega^2/\delta \omega_0^2}\;,
\end{equation}
where
\begin{equation}
\label{pumpbandwidths}
\delta q_0=2/w_0\;,\hspace{.6cm}\delta\omega_0=2/\tau_0
\end{equation}
denote the bandwidths of the pump
in the spatial frequency domain and in the temporal frequency domain respectively.

Let us now consider the limit of the PWPA approximation,
in which $w_0$ and $\tau_0$ tend to infinity and
\begin{equation}
A_0(z,\vec{q},\Omega)\rightarrow (2\pi)^{3/2}A_p~\delta(\vec{q})\delta(\Omega)\;.
\end{equation}
Under this condition Eqs. (\ref{waveq}) couple only pairs of phase-conjugated
modes $(\vec{q},\Omega)_1$ and $(-\vec{q},-\Omega)_2$ and can be solved analytically.
The unitary input-output transformations relating the field operators
at the output face of the crystal $a_j^{out}(\vec{q},\Omega)\equiv a_j(z=l_c,\vec{q},\Omega)$ to those at the
input face $a_j^{in}(\vec{q},\Omega)\equiv a_j(z=0,\vec{q},\Omega)$ take the following form
\begin{eqnarray}
\label{inputoutput}
a_1^{out}(\vec{q},\Omega)&=&U_1(\vec{q},\Omega)a_1^{in}(\vec{q},\Omega)
+V_1(\vec{q},\Omega)a_2^{in~\dag}(-\vec{q},-\Omega)\;,\\
a_2^{out}(\vec{q},\Omega)&=&U_2(\vec{q},\Omega)a_2^{in}(\vec{q},\Omega)
+V_2(\vec{q},\Omega)a_1^{in~\dag}(-\vec{q},-\Omega)\;,\nonumber
\end{eqnarray}
with
\begin{eqnarray}
\label{uv}
     U_1(\vec{q},\Omega)&=&
     \exp\left[i\frac{\delta_1(\vec{q},\Omega)-\delta_2(-\vec{q},-\Omega)-\Delta_0}{2}l_c\right]
     \left[\cosh(\Gamma(\vec{q},\Omega)l_c)
     +i\frac{\Delta(\vec{q},\Omega)}{2\Gamma(\vec{q},\Omega)}
     \sinh(\Gamma(\vec{q},\Omega)l_c)\right]\;,\\
     V_1(\vec{q},\Omega)&=&
     \exp\left[i\frac{\delta_1(\vec{q},\Omega)-\delta_2(-\vec{q},-\Omega)-\Delta_0}{2}l_c\right]
     \frac{\sigma_p}{\Gamma(\vec{q},\Omega)}
     \sinh(\Gamma(\vec{q},\Omega)l_c)\;,\nonumber\\
     U_2(\vec{q},\Omega)&=&
     \exp\left[i\frac{\delta_2(\vec{q},\Omega)-\delta_1(-\vec{q},-\Omega)-\Delta_0}{2}l_c\right]
     \left[\cosh(\Gamma(-\vec{q},-\Omega)l_c)
     +i\frac{\Delta(-\vec{q},-\Omega)}{2\Gamma(-\vec{q},-\Omega)}
     \sinh(\Gamma(-\vec{q},-\Omega)l_c)\right]\;,\nonumber\\
     V_2(\vec{q},\Omega)&=&
     \exp\left[i\frac{\delta_2(\vec{q},\Omega)-\delta_1(-\vec{q},-\Omega)-\Delta_0}{2}l_c\right]
     \frac{\sigma_p}{\Gamma(-\vec{q},-\Omega)}
     \sinh(\Gamma(-\vec{q},-\Omega)l_c)\;,\nonumber
\end{eqnarray}
and
\begin{mathletters}
\label{Gamma}
\begin{eqnarray}
      & &\Gamma(\vec{q},\Omega)
      =\sqrt{\sigma_p^2-\frac{\Delta(\vec{q},\Omega)^2}{4}}\;,\\
      & &\Delta(\vec{q},\Omega)
      =\Delta_0+\delta_1(\vec{q},\Omega)+\delta_2(-\vec{q},-\Omega)
      \approx k_{1z}(\vec{q},\Omega)+k_{2z}(-\vec{q},-\Omega)-k_0\;,\\
      & &\sigma_p=\sigma A_p\;.
\end{eqnarray}
\end{mathletters}
It is important to note that the gain functions $U_j$ and $V_j$ given by Eq.~(\ref{uv})
satisfy the following unitarity conditions:
\begin{mathletters}
\label{unitarity}
\begin{eqnarray}
     &&|U_j(\vec{q},\Omega)|^2-|V_j(\vec{q},\Omega)|^2=1\hspace{1cm}(j=1,2)\label{unit1}\\
     &&U_1(\vec{q},\Omega)V_2(-\vec{q},-\Omega)=
     U_2(-\vec{q},-\Omega)V_1(\vec{q},\Omega)\;,\label{unit2}
\end{eqnarray}
\end{mathletters}
which guarantee the conservation of the free-field commutation relations (\ref{commut})
after propagation.

\section{Mean intensity distribution}\label{meanint}
In the following we shall consider measurements either in the near
field or in the far field zones of the nonlinear crystal.
In order to simplify the notation we shall omit the explicit dependence of
the fields on the $z$ coordinate: when specification
is explicitly needed, the measured quantities will be labelled with $\pi$ or $\pi'$,
which will denote the near field and the far field detection planes respectively
(see scheme of Fig.\ref{fig1}a).
The analytical results given here and in the next sections are all
obtained within the PWPA; on the one hand
they generalize those illustrated in \cite{fluor} for a type I
crystal at degeneracy to a type II phase-matching configuration,
on the other hand they provide a good starting point to interpret
the results of the numerical model that includes the pulse shape and the finite
cross section of
of the pump beam. A more general input-output formalism which goes
beyond the PWPA is developed in appendix \ref{appendixA}.

With a stationary and plane-wave pump the near field intensity distribution
in the output plane of the crystal
clearly does not depend on $\vec{x}$ and $t$, because of the system invariance
with respect to translation in time and in the transverse plane.
Using input-output relations (\ref{inputoutput}) and recalling that the input fields is
are the vacuum states, we obtain easily
\begin{equation}
\label{nearint}
\langle I_j(\vec{x},t)\rangle_{\pi}
=\int\frac{d\Omega}{2\pi}
   \int\frac{d\vec{q}}{(2\pi)^2}
|V_j(\vec{q},\Omega)|^2\;,\hspace{.5cm}(j=1,2)\;.
\end{equation}
The function $|V_j(\vec{q},\Omega)|^2$ gives
the contribution of mode $(\vec{q},\Omega)_j$ to the total photon flux of beam $j$,
and is usually referred to as its spectral gain.
On the other hand, in the far field plane $\pi'$ the spatial Fourier modes are resolved spatially
and the photon distribution reflects the $\vec{q}$-dependence of these spectral functions.
From the expression of $\Gamma(\vec{q},\Omega)$ given by Eqs.~(\ref{Gamma}), we
see that down-conversion occurs most efficiently for the modes
satisfying the condition $\Delta(\vec{q},\Omega)<2\sigma_p$.
Using Eqs.~(\ref{Gamma}b) and (\ref{kz}),
the phase-mismatch accumulated during propagation
can be written in the form
\begin{equation}
     \label{mismatch}
     \Delta(\vec{q},\Omega)l_c=\Delta_0 l_c
     +{\rm sign}[k_1'-k_2']\frac{\Omega}{\Omega_0'}
     +\frac{\Omega^2}{\Omega_0''^2}
     -\rho_2 q_y
     -\frac{q_x^2+q_y^2}{q_0^2}
\end{equation}
where we assumed that the signal wave is ordinarily polarized, so that $\rho_1=0$,
and we introduced the parameters
\begin{equation}
\label{bandwidth}
     q_0=\sqrt{\frac{\overline{k}}{l_c}}\;,\hspace{1cm}
     \Omega_0'=\frac{1}{|k_1'-k_2'|l_c}\;,\hspace{1cm}
     \Omega_0''=\sqrt{\frac{2}{(k_1''+k_2'')l_c}}\;.
\end{equation}
where $\overline{k}=2k_1 k_2/(k_1+k_2)$.
They determine the characteristic bandwidths of SPDC both in the temporal frequency domain
and in spatial frequency domain.
In the type I phase-matching configuration we will consider in Sec.~\ref{numtypeI},
both the signal and the idler waves are ordinarily
polarized and are observed close to degeneracy, i.e. for $\omega_1=\omega_2=\omega_0/2$.
In this special case the temporal bandwidth is determined by
$\Omega_0\equiv\Omega_0''=\sqrt{1/|k_1''l_c|}$, since
$k_1(\omega)=k_2(\omega)$ implies that $\Omega_0'=\infty$.
Far from frequency degeneracy the emission
spectrum has a much narrower bandwidth, on the order of
$\Omega_0\equiv\Omega_0'$ which is about 2 to 3 orders of magnitude smaller than
$\Omega_0''$ (for a typical crystal length of few millimeters).
On the other hand, in type II crystals the signal and idler waves
are characterized by different polarizations and frequency dispersion relations, so that
$\Omega_0\equiv\Omega_0'$ remains finite even for $\omega_1=\omega_2$.
The spatial bandwidth $q_0$ gives
the range of transverse wave-vectors for which the gain spectrum $|V_j(\vec{q},\Omega)|^2$
is close to its maximum value, $\sinh^2[\sigma_p l_c]$,
when the $j^{th}$ field is observed at a given frequency $\omega_j+\Omega$
(the gray region shown schematically in Figs.\ref{fig1}~b and c for $\Omega=0$).
We remark that $1/\rho_2 l_c$ and $q_0$ are about the same order of magnitude
as long as $l_c$ remains in the millimeter range.

For definiteness we assume that the far field is observed in the focal plane of a thin lens
of focal length $f$ which performs the Fourier transformation of the field from
the output face of the crystal (the so-called $f-f$ system).
The field operators in the focal plane $\pi'$ at $z=l_c+2f$ (see Fig.\ref{fig1}a),
which we denote with $b_{1,2}(\vec{x},t)$, are related to
those in the output plane of the crystal by the following Fresnel transformation
\begin{mathletters}
\label{farfield}
\begin{eqnarray}
b_j(\vec{x},t) &=&\int d\vec{x}~'h_j(\vec{x},\vec{x}~') a_j^{out}(\vec{x}~',t)\;,\label{farfielda}\\
h_j(\vec{x},\vec{x}~')&=&\frac{-i}{\lambda_j f}
e^{-\frac{2\pi i}{\lambda_j f}\vec{x}\cdot\vec{x}~'}\;,
\hspace{1cm}(j=1,2)\;,\label{farfieldb}
\end{eqnarray}
\end{mathletters}
where $\lambda_j=2\pi c/\omega_j$, (j=1,2),
are the free-space wavelengths corresponding to the carrier
frequencies. Using the input-output relations (\ref{inputoutput})
and unitarity relations (\ref{unitarity}) we can evaluate
the mean intensity distribution of the two fields with the following approximate
expression
\begin{equation}
\label{farint}
\langle I_j(\vec{x},t) \rangle_{\pi'}
\approx\frac{1}{S^{(j)}_{diff}}\int\frac{d\Omega}{2\pi}|\overline{V}_j(\vec{x},\Omega)|^2\;,
\hspace{1cm}(j=1,2)\;.
\end{equation}
where we introduced the barred gain functions defined in real space
\begin{equation}
\label{uvbarrati} \overline{U}_j(\vec{x},\Omega)
=U_j\left(\frac{2\pi}{\lambda_j f}\vec{x},\Omega
\right)\:,
~~~~~~~~
\overline{V}_j(\vec{x},\Omega)
=V_j\left(\frac{2\pi}{\lambda_j f}\vec{x},\Omega
\right)\:,~~~~~(j=1,2),
\end{equation}
and $S^{(j)}_{diff}=(\lambda_{j}f)^2/S_A$, (j=1,2), denotes the resolution areas
in the far field plane at the S/I wavelengths, with $S_A$
being the area characterizing the dimension of the system in the transverse plane.
As it is shown in \cite{fluor}, Eq.(\ref{farint}) can be obtained
by assuming that a pupil of area $S_A\gg 1/q_0^2$ is put on the crystal exit face.
Assuming that the transverse dimensions of the
crystal are large compared to the pump waist, $S_A$ can be identified
with the effective cross section area of the pump beam.
Eq.(\ref{farint}) represents a good approximation provided the pump beam shape changes negligibly
during propagation in the crystal and behaves therefore as a plane-wave. This happens when the Rayleigh length
characterizing the Gaussian pump beam divergence, $z^0_R=\pi w_0^2/\lambda_0$, and its analogue
characterizing dispersion, $z^0_{disp}=\tau_0^2/2k_0''$, are much
longer than the crystal length $l_c$. The same conditions can also be written in terms
of the pump spatial and temporal bandwidths (\ref{pumpbandwidths}) as
\begin{equation}
\label{largepump}
\frac{\delta q_0}{q_0}\ll 1\;,
\hspace{1cm}
\frac{\delta \omega_0}{\Omega_0}\ll 1\;.
\end{equation}
At the considered carrier frequencies $\omega_1$ and $\omega_2$
the gain functions $V_j(\vec{q},\Omega=0)$ are maximal and perfect phase-matching is achieved when
the equations $\Delta(\pm \vec{q},\Omega=0)=0$ are satisfied, with the plus sign
for field $1$, and the minus sign for field $2$ (see Eqs.(\ref{uv}) and (\ref{Gamma}b)).
More explicitly, as can be seen using expression (\ref{mismatch}), they can be written as
\begin{equation}
\label{rings}
\frac{q_x^2}{q_0^2}+\left(\frac{q_y}{q_0}\pm\frac{1}{2}\rho_2 l_c q_0\right)^2
=\Delta_0 l_c+\left(\frac{1}{2}\rho_2 l_c q_0\right)^2\;.
\end{equation}
Provided that $\Delta_0 l_c>-\frac{1}{4}\rho_2^2 l_c^2 q_0^2$, we have therefore two
circles of radius $q_R$ and centered at $(q_x=0,q_y=\pm q_C)$,
with
\begin{mathletters}
\label{qRqC}
\begin{eqnarray}
q_C&=&\frac{1}{2}\rho_2 l_c q_0^2
=\frac{1}{2}\overline{k}\rho_2\;,\label{qC}\\
q_R&=&q_0\sqrt{\Delta_0 l_c+\frac{q_C^2}{q_0^2}}
=\sqrt{\overline{k}\Delta_0+\frac{1}{4}(\overline{k}\rho_2)^2}\label{qR}\;.
\end{eqnarray}
\end{mathletters}
They are plotted in Fig.\ref{fig1}b,c, respectively for a type II and a type I phase-matching
configuration. Modes close to these circles within $q_0$ (gray annuli in the figure)
give a non negligible contribution to the down-converted field.
In the detection plane they give rise to characteristic couples of rings
which have been observed in many experiments on spontaneous
parametric down-conversion (see e.g.~\cite{lantz2,ditrapani,jost}). It should be stressed
that, without any spectral filtering, emission occurs on a very wide
range of wavelengths and emission angles, as allowed by the phase-matching conditions (see e.g. \cite{ditrapani}).
However, from an experimental point of view, a particular couple of rings can always be selected with the use of
frequency filters centered at the chosen frequencies $\omega_1$ and $\omega_2=\omega_0-\omega_1$.
Noting that there is the following mapping between the spatial frequency plane and the
far field plane:
\begin{eqnarray}
\label{qxspace}
& &(q_x,q_y)\rightarrow\frac{\lambda_1 f}{2\pi}(q_x,q_y) \hspace{.5cm}{\rm for}\;{\rm field}\;1,\\
& &(q_x,q_y)\rightarrow\frac{\lambda_2 f}{2\pi}(q_x,q_y) \hspace{.5cm}{\rm for}\;{\rm field}\;2,\nonumber
\end{eqnarray}
it is easily seen that the ring radii, $x_R^{(1,2)}=\frac{\lambda_{1,2} f}{2\pi}q_R$,
and their distance from the $z$-axis, $y^{(1,2)}_C=\frac{\lambda_{1,2} f}{2\pi}q_C$,
are generally different except when observation is performed at frequency degeneracy
(i.e. for $\lambda_1=\lambda_2$).
In a type I phase-matching configuration these rings are concentric,
since there is no spatial walk-off between the two fields ($\rho_2=0$)
and the radial symmetry of the system is preserved.

Fig.~\ref{fig2} and \ref{fig3} illustrate the kind of far field
patterns that can be obtained at frequency degeneracy in a type I and a type II crystal,
respectively, assuming the pump field has a large beam waist (in
practice, the condition $\delta q_0\ll q_0$ must be fulfilled).
They are obtained by numerical
integration of the classical field equations (\ref{qwaveq}), with a
white input noise which simulates
the vacuum fluctuations that trigger the process,
as it will be described in the next sections.
These 2-D simulations do not include the temporal dimension and
cannot be used to obtain quantitative results; however, they
provide some insight on the spatial features of the far field patterns
generated in single pump shots, as they could be
observed experimentally
by using a narrow band filter at $\Omega=0$.

In the type I phase-matching case, being at frequency degeneracy,
the two rings merge into one that contains both the signal and idler modes.
It should be noted that $q_R$ and $q_C$
do not depend on the crystal length whereas $q_0$ scales as $1/\sqrt{l_c}$,
so that the thickness of the rings at a given temporal frequency is larger for a shorter crystal.

\section{Near and far field correlations}
\label{correlations}
We now define explicitly the quantities that can be measured in an experiment
in order to put in evidence the S/I
correlations in the spatial domain we are investigating.
We assume that the signal and idler beams are spatially separated in the detection
plane and are measured
over two detection areas which we denote with $R_1$ and $R_2$.

In the far field, correlations find their origin in the conservation
of the transverse momentum of the generated photon pairs.
Therefore, in order to find maximal correlation, $R_1$ and $R_2$ must
correspond couples of phase-conjugate
modes, such as those indicated with the black squares in Fig.~\ref{fig1}~b,c.
For simplicity, in order to avoid the heavy notations which arise
if $\lambda_1\neq\lambda_2$, we shall restrict our analysis to
the frequency degenerate case, indicating with $\lambda$ both
$\lambda_1$ and $\lambda_2$. Phase-conjugate modes
are then mapped by the lens into symmetrical points in plane $\pi'$
according to relation (\ref{qxspace}) and $R_1$ and $R_2$ must be taken symmetrical.

On the other side, near field correlations
arising from the position entanglement of the twin photons
are expected to be observed if $R_1$ and $R_2$ occupy the same
region of the near field plane. In practice a type II phase-matching
configuration should be considered, so that the use of a polarizing beam-splitter
and lens systems allows the imaging of the S/I near fields
on two physically separated detection planes (see detection scheme
illustrated in Fig.\ref{fig4}).

If the detectors are the pixel of a CCD camera, as in the experiment
described in \cite{ditrapani}, they do not allow any spectral
measurement due to the very low resolution power of the device in the time domain.
They simply measure the total number of incoming photons
down-converted in each single pump shot and the measurement time $T_d$ can be identified
with the pump pulse duration.
We introduce therefore the operators corresponding to the number of photons
collected by the two detectors in the finite time window $[-T_d/2,T_d/2]$
\begin{equation}
     N_j=\int_{R_j}d\vec{x}\int_{-T_d/2}^{T_d/2}dt~I_j(\vec{x},t)\;\hspace{.5cm}(j=1,2),
\end{equation}
The measurable quantity which is capable of displaying
the quantum nature of the photon number statistics in the spatial domain is the variance of the
photon number difference, $N_-=N_1-N_2$, which can be written in the form
\begin{eqnarray}
\label{variance}
\langle(\delta N_-)^2 \rangle
=\langle N_+\rangle
  +\langle:(\delta N_1)^2:\rangle
  +\langle:(\delta N_2)^2:\rangle
  -2\langle \delta N_1 \delta N_2 \rangle\;.
\end{eqnarray}
$\delta N_j=N_j-\langle N_j \rangle$ and $\delta N_-=N_{-}-\langle N_- \rangle$
denote the photon number fluctuation operators associated to $N_j$ and $N_-$ and
the colon ":" denotes normal ordering (n.o.) for the expectation values.
In Eq.(\ref{variance}) the shot noise contribution, i.e. the total number of photons
intercepted by the two detector $\langle N_+\rangle=\langle N_1\rangle+\langle N_2\rangle$,
has been explicitly separated from the term which describes the field correlations.
We define
\begin{equation}
\label{dnij}
\langle:\delta N_i \delta N_j:\rangle
= \int_{R_i}d\vec{x}
 \int_{R_j}d\vec{x}~'
 \int_{-T_d/2}^{T_d/2}dt
 \int_{-T_d/2}^{T_d/2}dt'
G_{ij}(\vec{x},t,\vec{x}~',t')\;,
\hspace{.5cm}(i,j=1,2)\;,
\end{equation}
where
\begin{equation}
\label{correl}
G_{ij}(\vec{x},t,\vec{x}~',t')
=\langle:I_i(\vec{x},t) I_j(\vec{x}~',t'): \rangle
-\langle
I_i(\vec{x},t)\rangle \langle I_j(\vec{x}~',t')
\rangle\;,
\hspace{.5cm} (i,j=1,2)\,.
\end{equation}
are the n.o. self- and cross-photon number correlation functions of the S/I beams.
Notice that in the non-degenerate case (type II or type I far from
frequency degeneracy) we have
$\langle :I_1(\vec{x},t) I_2(\vec{x}~',t'):\rangle=
\langle I_1(\vec{x},t) I_2(\vec{x}~',t')\rangle$.

We now focus on the analytical results that can be deduced from the PWPA.
We shall consider explicitly only the case of type II phase-matching. The case of type I,
at least in the far field, can be described with a similar treatment and has been
already discussed in \cite{fluor}.
Assuming that the detection time $T_d$ is large compared to the coherence
time $\tau_{coh}=\Omega_0^{-1}$, as it is usually the case,
we have
\begin{equation}
\langle :\delta N_i \delta N_j :\rangle
\approx T_d
\int_{R_i}d\vec{x}
\int_{R_j}d\vec{x}'
~G_{ij}(\vec{x},\vec{x}\,',\Omega=0)
\hspace{.5cm} (i,j=1,2)\,.
\end{equation}
where $G_{ij}(\vec{x},\vec{x}\,',\Omega=0)$ is the Fourier transform of the function (\ref{correl})
with respect to $t-t'$ (notice that in a cw regime this function depends only on $t-t'$).
Next, the Gaussian character of the field statistics
allows to express fourth order correlations in terms of second-order
correlations (see App.\ref{appendixA}, Eq. (\ref{corrb}))
In this way the photon number correlation defined by Eq.~(\ref{dnij}) in a plane $z$ can
be written as
the photon number correlations defined by Eq.(\ref{dnij}) in a
plane $z$ can then be written as
\begin{equation}
\label{dnij2}
\langle:\delta N_i \delta N_j: \rangle_z=T_d
\int_{R_i}d\vec{x}
\int_{R_j}d\vec{x}~'
\int\frac{d\Omega}{2\pi}|\Gamma_{ij}^{(z)}(\vec{x},\vec{x}~',\Omega)|^2\;,
\hspace{.5cm}(i,j=1,2)\;,
\end{equation}
where
\begin{mathletters}
\begin{eqnarray}
\label{g11z}
\Gamma_{jj}^{(z)}(\vec{x},\vec{x}~',\Omega)
&=&\int d\tau e^{-i\Omega\tau}
\langle a_j^{\dag}(z,\vec{x},t+\tau)a_j(z,\vec{x}~',t)\rangle\;,
\hspace{.5cm}(j=1,2)\;\\
\Gamma_{12}^{(z)}(\vec{x},\vec{x}~',\Omega)
&=&\int d\tau e^{-i\Omega \tau}
\langle a_1(z,\vec{x},t+\tau)a_2(z,\vec{x}~',t)\rangle\;,
\label{g12z}
\end{eqnarray}
\end{mathletters}
are the only correlation spectra of the S/I fields which do not vanish
when the input field is in the vacuum state. We assumed here implicitly that
propagation in free space occurs without losses.

In the far field plane $\pi'$ the self and cross correlation functions are given by
\begin{mathletters}
\label{gijfar}
\begin{eqnarray}
\label{gijfarA}
\Gamma_{jj}^{(\pi')}(\vec{x},\vec{x}~',\Omega)&=&\delta(\vec{x}-\vec{x}~')
                                    |\overline{V}_j(\vec{x},\Omega)|^2\;,~~~~~~(j=1,2),\\
\label{gijfarB}
\Gamma_{12}^{(\pi')}(\vec{x},\vec{x}~',\Omega)&=&-\delta(\vec{x}+\vec{x}~')
                        \overline{U}_1(\vec{x},\Omega)\overline{V}_2(-\vec{x},-\Omega)\;.
\end{eqnarray}
\end{mathletters}
as it can be inferred using the Fresnel transformations (\ref{farfield}).
In this case, both correlation functions display a delta-like peak, located at $\vec{x}~'=\vec{x}$
for the self correlation, and at $\vec{x}~'=-\vec{x}$ for the cross correlation.
The delta-like character of the correlations derives from the unphysical assumption
that the transverse dimensions of the system are infinite.
In \cite{fluor} which deals
only with far field correlations the transverse size of the system were
taken into account {\it a posteriori}
by considering a finite aperture $S_A$ placed at the crystal output face, assuming a condition
equivalent to (\ref{largepump}) is met.
With this approach, we found that far field correlations
are localized within the resolution area determined by the size of this aperture
(that is $S_{diff}=(\lambda f)^2/S_A$ with the $f-f$ lens system).
This procedure also eliminates the cumbersome divergencies
arising from the singularity of the spatial delta-functions,
allowing the formal substitution $\delta(\vec{x}=0)\rightarrow 1/S_{diff}$
as has been done in Eq.(\ref{farint}).
By integrating over two symmetric detection pixels of area much larger than
this resolution length we obtain
\begin{mathletters}
\begin{eqnarray}
& &\langle :(\delta N_1)^2:\rangle_{\pi'}
  =\langle :(\delta N_2)^2:\rangle_{\pi'}
  =\frac{1}{S_{diff}}
\int\frac{d\Omega}{2\pi}
\int_{R_1}d\vec{x}~
|\overline{V}_1(\vec{x},\Omega)|^4\;,\\
& &\langle \delta N_1 \delta N_2 \rangle_{\pi'}
=\frac{1}{S_{diff}}
\int\frac{d\Omega}{2\pi}\int_{R_1}d\vec{x}~
|\overline{U}_1(\vec{x},\Omega)\overline{V}_2(-\vec{x},-\Omega)|^2\;.
\end{eqnarray}
\end{mathletters}
In order to obtain these expressions, both unitarity relations (\ref{unit1})
and (\ref{unit2}) must be used, together with the fact that the integration
area $R_1$ and $R_2$ are taken symmetric with respect to the origin.
We then easily get
\begin{mathletters}
\label{noisered}
\begin{eqnarray}
\langle :(\delta N_1)^2:\rangle_{\pi'}
+\langle :(\delta N_2)^2:\rangle_{\pi'}
-2\langle \delta N_1 \delta N_2 \rangle_{\pi'}
&=&
-\frac{1}{S_{diff}}
\int\frac{d\Omega}{2\pi}
\int_{R_1}d\vec{x}
~|\overline{V}_1(\vec{x},\Omega)|^2\;,\\
&=&-\langle N_+\rangle_{\pi'}\;,
\end{eqnarray}
\end{mathletters}
which implies $\langle (\delta N_-)^2 \rangle_{\pi'}=0$, as it follows from Eq.~(\ref{variance}).
A more rigorous
approach to the issue concerning the finite resolution of the system is given
in appendix \ref{appendixB}, where we derive an approximate solution of the propagation equations
which include the finite pump dimensions in the limit (\ref{largepump}).
It is shown that the width of the far field correlation peaks are
indeed on the order of $x_{diff}=(\lambda f/2\pi)\delta q_0$,
the resolution length determined by the pump beam waist $w_0$.

In the near field plane $\pi$ (at $z=l_c$), using
Eqs.(\ref{inputoutput}) and (\ref{commut}) we obtain the expressions
\begin{mathletters}
\label{gijnear}
\begin{eqnarray}
\Gamma_{jj}^{(\pi)}(\vec{x},\vec{x}~',\Omega)
&=&\int\frac{d\vec{q}}{(2\pi)^2}
e^{-i\vec{q}\cdot(\vec{x}-\vec{x}~')}
|V_j(\vec{q},\Omega)|^2\;,~~~~~(j=1,2),\label{g11}\\
\Gamma_{12}^{(\pi)}(\vec{x},\vec{x}~',\Omega)
&=&\int\frac{d\vec{q}}{(2\pi)^2}
e^{i\vec{q}\cdot(\vec{x}-\vec{x}~')}
U_1(\vec{q},\Omega)V_2(-\vec{q},-\Omega)\;,\label{g12}
\end{eqnarray}
\end{mathletters}
which depend only on the offset between the two points, $\vec{x}-\vec{x}~'$,
as a consequence of the invariance of the system with respect to translations in the
transverse plane which follows from the PWPA.
Provided that the typical scale of variation of the function appearing under the
integrals in Eqs.~({\ref{g11}) and (\ref{g12}) is $q_0$, we expect that these
correlations are localized in a region of size
\begin{equation}
\label{xcoh}
x_{coh}\equiv 1/q_0\approx\sqrt{l_c/\overline{k}}\;,
\end{equation}
a quantity which can be identified with the transverse coherence length of the
down-converted fields. This finite correlation length comes from
the spread out of the generated photons due to diffraction,
which increases proportionally to the square root of the propagation distance;
we can therefore expect that the detection areas must be
larger than this coherence area in order to
measure good correlations in the near field.

We incidentally note that the cross correlation function $\Gamma_{12}$ displays a localized peak,
that in the near field is located at $\vec{x}~'=\vec{x}$ (Eq.(\ref{g12})),
while in the far field is located at $\vec{x}~'=-\vec{x}$ (Eq.(\ref{gijfarB})).
The $\vec{x}\leftrightarrow\vec{x}~'$ correlation in the near field reflects
the entanglement in position of the twin photons,
while the $\vec{x}\leftrightarrow-\vec{x}~'$ correlation in the far field comes
from their entanglement in momentum.

For what concerns the near field, let us consider more in general what happens in a generic
plane of coordinate $z$ close to the crystal output face. As it is shown schematically
in Fig.\ref{fig4},
we consider a measurement in which
the signal and the idler fields are separated with a polarizing beam splitter placed beyond
the crystal. The two lenses $L$ and $L'$ put in the signal and idler arms performs the imaging
of plane $z$ onto two distinct detection planes (note that we have here two $2f-2f$ lens systems,
while in the case of the far field measurement considered previously we had a single $f-f$ system).
The S/I photons are collected by means of two square pixel detectors
$R_1$ and $R_2$, centered at the positions $\vec{x}_1$ and $\vec{x}_2$ respectively.
The propagation from the crystal exit face $z=l_c$ to the detection planes can
be described by a Fresnel transformation of the form (\ref{farfielda}) with the kernel
\begin{equation}
h(\vec{x},\vec{x}~')=\frac{-i}{\lambda(z-l_c)}e^{\frac{-i}{\lambda(z-l_c)}|\vec{x}-\vec{x}~'|^2}
\end{equation}
where inessential phase factors due to the presence of the lenses have been omitted.
By using this transformation
inside Eqs.(\ref{g11z}) and (\ref{g12z})
we can calculate explicitly
each term on the r.h.s. of Eq.(\ref{variance}).
By performing explicitly
the integration over the square pixel areas in Eq.(\ref{dnij2}), we obtain
\begin{mathletters}
\label{cor}
\begin{eqnarray}
\label{cor1}
& &\langle :(\delta N_j)^2:\rangle_{z}
=T_d
\int d\vec{q}
\int d\vec{q}~'
H_{11}(\vec{q},\vec{q}~')
\int\frac{d\Omega}{2\pi}|V_j(\vec{q},\Omega)|^2|V_j(\vec{q}~',\Omega)|^2\;,\hspace{.5cm}(j=1,2)\\
\label{cor2}
& &\langle \delta N_1 \delta N_2 \rangle_{z}
=T_d
\int d\vec{q}
\int d\vec{q}~'
H_{12}(\vec{q},\vec{q}~')
\int\frac{d\Omega}{2\pi}\:\overline{U}_1(\vec{q},\Omega)\overline{V}_2(-\vec{q},-\Omega)
                          \overline{U}_1^*(\vec{q}~',\Omega)\overline{V}_2^*(-\vec{q}~',-\Omega)\;,\\
& &\langle N_+ \rangle_{z} =2 T_d \:d^2\int \frac{d\vec{q}}{(2\pi)^2}
                   \int\frac{d\Omega}{2\pi} |V_j(\vec{q},\Omega)|^2
\end{eqnarray}
\end{mathletters}
where the functions $H_{ij}$ depend on the square pixel size $d$ and on their relative
positions through the relations
\begin{mathletters}
\label{HIJ}
\begin{eqnarray}
\label{H11}
H_{11}(\vec{q},\vec{q}~')
&=&
(d/2\pi)^4
~\mbox{sinc}^2
\left[\frac{(q_x-q_x')d}{2}\right]
~\mbox{sinc}^2
\left[\frac{(q_y-q_y')d}{2}\right]\;,\\
H_{12}(\vec{q},\vec{q}~')
&=&e^{-i\frac{\lambda (z-l_c)}{2\pi}(q^2-q'^2)+i(\vec{q}-\vec{q}~')\cdot(\vec{x}_1-\vec{x}_2)}
H_{11}(\vec{q},\vec{q}~').
\label{H12}
\end{eqnarray}
\end{mathletters}
We first note that if the detection areas are reduced well below the coherence area $x_{coh}^2$, the fluctuations
of $N_-$ approach shot noise. Indeed, in the limit $d\ll x_{coh}$
we can replace $H_{11}(\vec{q},\vec{q}~')$ with
$H_{11}(\vec{q},\vec{q})=d^4/(2\pi)^4$ in Eqs.(\ref{cor1}) and (\ref{cor2}), from which it can be verified
that the correlation terms scale
as $d^4/x_{coh}^4$ while the shot noise contribution scale as $d^2/x_{coh}^2$
(to evaluate this scaling it should
be noted that for a fixed $\Omega$, the area in $\vec{q}$-space where the gain functions are not
negligible is on the order of $q_0^2=1/x_{coh}^2$).
On the other hand, if the detection areas are large enough
with respect to $x_{coh}^2$, the substitutions
\begin{equation}
\label{Hdelta}
H_{ij}(\vec{q},\vec{q}~')\rightarrow
\frac{d^2}{(2\pi)^2}\delta(\vec{q}-\vec{q}~')
\end{equation}
can be used for evaluating both (\ref{cor1}) and (\ref{cor2}); this leads
to vanishing fluctuations in the measurement of $N_-$ as for the
far field case (again, unitarity relations (\ref{unitarity}) must be used in order
to obtain this result).
However the condition "large enough"
is  more stringent
for cross-correlation than for self-correlation.
Let us first consider the case $z=l_c$, and $\vec{x}_1=\vec{x}_2$.
The function appearing under the integral in (\ref{cor1}) is always positive. By contrast,
the function appearing under the integral in (\ref{cor2}) is an oscillating function,
which becomes positive only for
an infinite pixel size, when the limit (\ref{Hdelta})
is strictly achieved. This feature tends to lower
$\langle \delta N_1 \delta N_2 \rangle_{z}$
with respect to $\langle :( \delta N_1)^2:\rangle_{z}$ and $\langle :(\delta N_2)^2:\rangle_{z}$.
As a consequence of this behavior the fluctuations of $N_-$ will therefore exceed shot noise,
as it can be easily inferred from expression (\ref{variance}).
As a matter of fact, as we shall see in Sec.\ref{numtypeIIn},
$\vec{x}_1 = \vec{x}_2$ and $z =l_c$ are not the better choices
to minimize the fluctuations of $N_-$, and
special care in the positioning of
the detectors is necessary in order to compensate both the effect of
diffraction and spatial walk-off between the S/I fields, which  are included
by the phase of the function
$U_1(\vec{q},\Omega)V_2(-\vec{q},-\Omega)$.

Finally we note that if the losses of the detection process
are taken into account, the ideal result $\langle (\delta N_-)^2\rangle=0$
must be replaced with
\begin{equation}
\langle (\delta N_-)^2\rangle=\eta(1-\eta)\langle N_+\rangle\;,
\end{equation}
$\eta$ denoting the finite quantum efficiency of the detectors.

\section{Numerical results}
\label{nummodel}
We now present the results obtained from the numerical model that
includes the effects of the finite pump. The quantum averages in
which we are interested (i.e. mean  photon numbers and photon number
correlations) are evaluated through a stochastic method
based on the Wigner representation. With respect to other representations
in phase space, the Wigner representation presents the advantage that
the c-number stochastic equations equivalent
to the equations for the field operators (\ref{qwaveq})
do not contain Langevin noise terms (because of linearity and absence of dissipation)
and are therefore identical to the classical propagation equations (\ref{waveq}).
The statistical character of the quantum fields is wholly contained in the
stochastic input field.
We therefore proceed as follow:

1) We generate the input field with the
appropriate phase-space probability distribution, which is a
Gaussian white noise with zero mean, corresponding to
the vacuum state in the Wigner representation \cite{random}.

2) We perform the numerical integration of Eqs.(\ref{waveq}). We use a
split-step algorithm \cite{numrec} which integrates separately the terms
describing linear propagation and
the term describing the wave
mixing process: the former are integrated in Fourier space, the
latter in real space.

3) The obtained output fields are used
to evaluate the correlation functions of interest.
The procedure must be reiterated a sufficiently large number of
times, so that the  stochastic averages performed
becomes good approximations to the corresponding quantum
expectation values. Furthermore,
some corrections are usually necessary in order to convert
them to the desired operator ordering
(the Wigner representation yields quantum expectation values of
symmetrized operator products).

\subsection{Far field correlation in a type I crystal at degeneracy}\label{numtypeI}
We first focus our attention on far field correlations that can be observed in a type I
crystal with emission close to degeneracy
(the case illustrated in Fig.\ref{fig3}).
Since there is no spatial and temporal walk-off between the signal and
the idler modes, the most significant parameters in play are the
spatial bandwidth $q_0=\sqrt{\overline{k}/l_c}$ and
the the ring radius $q_R=\sqrt{\overline{k}\Delta_0}$, which are determined by the crystal
length and the collinear phase-mismatch parameter (see
Eqs.(\ref{bandwidth}) and (\ref{qR})).
To a large extent, the analysis that we have performed
does not depend on the particular type I crystal that is considered.
We shall however refer to the specific case of LBO (LiB$_3$O$_5$: {\it
lithium triborate}) set in the type I configuration described
e.g. in \cite{lantz2,ditrapani}. The pump field operates at
$\lambda_0=532$ nm and propagates in the $XY$ crystal plane
forming an angle with the $X$-axis close to $11^o$, for which
collinear phase-matching at degeneracy is achieved. In order to
evaluate the characteristic bandwidths and walk-off parameters
given by Eq.~(\ref{bandwidth}), we have used the Sellmeier
dispersion relation coefficients which can be found in
\cite{handbook}.
Considering a crystal length of 5 mm, the coherence time is
$\tau_{coh}=\Omega_0^{-1}\approx 0.01$ ps. Assuming the pump
pulse has a duration of about 1.5 ps, as in the experiment described in
\cite{ditrapani}, the ratio $\delta \omega_0/\Omega_0$
is as small as $10^{-2}$.
The analytical results obtained within the plane-wave and cw pump
approximation are therefore expected to provide good insight as long as
the ratio $\delta q_0/q_0$ remains small compared to unity.
The temporal/spatial
walk-off between the signal and the pump field are negligible
unless the pump pulse time/beam waist are exceedingly small, a
situation we do not consider, since it is too far from the ideal plane
wave pump limit (it would prevent the observation of any spatial
quantum correlation effects).

Fig.~\ref{fig5} displays the intensity distribution obtained
numerically from a single integration of the propagation equations,
for increasing values of the
pump beam spatial bandwidth $\delta q_0=2/w_0$, both in the near field and in the
far field planes. As already mentioned, these simulations with two
transverse dimensions
do not include the temporal dimension and are only meant to
provide some insight on the spatial features of the
down-converted field generated in a single pump shot.
In the near field, the intensity follows the Gaussian
profile of the pump and displays a noisy spot pattern with a characteristic wavelength
$~\pi/q_R$.
In the far field, the intensity peaks (white spots in the figures) always appear in symmetrical
pairs as in the plane-wave pump case illustrated in Fig.\ref{fig3}~b.
However, they become broader and broader as the pump beam waist
$w_0$ beam is reduced. Their size in the observation plane $\pi'$ is on the order of
the resolution length imposed by the
finite transverse size of the pump beam waist, i.e.
$x_{diff}=\frac{\lambda f}{2\pi}\delta q_0$.
The following quantitative
evaluations of the amount of correlation include instead the temporal dimension but consider
only one transverse dimension in space, since a full 2-D
calculation (2 transverse dimensions + time) would have required an exceedingly
long CPU calculation time for our computer.

Some insight can be gained by looking at the n.o. photon-number correlation
$\langle :\delta N(\vec{x})\delta N(\vec{x}~'):\rangle_{\pi'}$
between two pixels centered at $\vec{x}$ and $\vec{x}~'$,
defined by an expression similar to Eq.(\ref{dnij}) which
refers to the type II phase-matching configuration. The vectors
$\vec{x}$ and $\vec{x}~'$ denote here the positions of the two pixel
detectors in the transverse plane (typically two pixels of a CCD camera), their size
being determined in the simulation by the spatial step of the numerical grid.
In the degenerate case considered here the indexes $i,j$ are dropped
since the signal and idler fields are not distinguishable.
Integration in time is performed over an interval $T_d$
which is taken larger than the pump pulse time, $\tau_0=1.5$ps, so that all the
down-converted photons generated by a single pump pulse are collected by the two detectors.
Its 1-D (one transverse dimension + time) numerical evaluation
is plotted in Fig.\ref{fig6} as a function of $x$, keeping $x'$ fixed
at $x'=(\lambda f/2\pi)q_R$, where the gain is maximum, and for
different values of the ratio $\delta q_0/q_0$.
The $x$-coordinate is normalized to $x_0=\frac{\lambda f}{2\pi}q_0$,
the spatial scale of the photon number distribution in the far field plane.
According to these simulations, the widths of the two peaks is on the order of $x_{diff}$,
the resolution length imposed by the pump beam transverse dimensions.
The correlation peak at $x=-x'$,
left in the plots of Fig.\ref{fig6}, is always higher than the
correlation peak on the right at $x=x'$.
In a similar way as it was shown in Sec.\ref{correlations} (see Eqs.(\ref{variance})
and Eqs.(\ref{noisered})),
this particular behavior of the n.o. correlation function indicates the possibility
that the fluctuations of $N_-=N_1(\vec{x})-N_2(\vec{x}\,')$ vanish when measured
from two symmetrical pixels (i.e. taking $\vec{x}=-\vec{x}\,'$).
Indeed, for two disconnected pixels we have
\begin{equation}
\langle (\delta N_-)^2 \rangle=\langle N_+ \rangle
+\langle :(\delta N(\vec{x}))^2:\rangle
+\langle :(\delta N(\vec{x}\,'))^2:\rangle
-2\langle \delta N(\vec{x}) \delta N(\vec{x}\,') \rangle
\end{equation}
where $\langle N_{+}\rangle\equiv\langle N(\vec{x})+ N(\vec{x}\,') \rangle$
represents the shot noise for $N_-$; therefore, since
$\langle (\delta N_-)^2 \rangle$
is always a non negative quantity, the following inequality holds:
\begin{equation}
\label{positiv}
2\langle \delta N(\vec{x}) \delta N(\vec{x}\,') \rangle
-\langle :(\delta N(\vec{x}))^2:\rangle
-\langle :(\delta N(\vec{x}\,'))^2:\rangle
\leq
\langle
N_+
\rangle\;.
\end{equation}
In particular, when the equality
sign holds in Eq.~(\ref{positiv}), the maximum amount of correlation
between the points $\vec{x}$ and $\vec{x}\,'$ is achieved, implying thus a complete suppression of the noise of $N_-$.

However the amount of noise in $N_-$ depends on the actual size of the detectors used to probe the correlation.
We have evaluated numerically the variance of $N_-$
for two symmetric detection areas, by varying the size of the detectors
as could be obtained in practice by grouping several pixels of a CCD.
The results are shown in Fig.\ref{fig7}, where the different lines
correspond to different values of the ratio $\delta q_0/q_0$, that is to different values
of the pump beam waist.
Fluctuations are well below shot noise when the detector size $d$
is larger than $x_{diff}$, that is for $d/x_0>\delta q_0/q_0$, which in turn
implies that the detection size must be larger than the width of the correlation peaks.
These simulations show that the localized character of
the correlation predicted by the plane-wave
pump theory is well preserved as long as the pump beam waist is not too small.
Only in the worst case
considered, with $\delta q_0/q_0=0.5$, the noise reduction factor
$\langle (N_-)^2 \rangle_{\pi'}/\langle N_+\rangle_{\pi'}$ is
is never close to zero unless the detectors cover the whole widths of the
the ring pattern. It should be noted that
a further increase of the ratio $\delta q_0/q_0$ would lead to single-mode emission:
the inverse of these ratio provides indeed an estimation of the number of spatial modes that are
efficiently amplified when the field is observed at a fixed temporal frequency.

\subsection{Far field correlation in type II crystals}
\label{numtypeIIf}
For the case of a type II phase-matching configuration, we shall consider explicitly the system
described in \cite{ditrapani}:
a $~1.5$ps high intensity laser pulse is injected in a 4 mm long beta barium borate (BBO) crystal cut
for type II phase-matching. In the example we consider the pump is oriented at an angle close
to $48.2^o$ with respect to the crystal axis and SPDC is observed around the
degenerate wavelength $\lambda_1=\lambda_2=704nm$ with a $10nm$
interference filter. For the chosen parameters, the radii of the two rings $q_R$ vanish,
which is the situation illustrated in Fig.\ref{fig2}c. As for the type I configuration,
we investigated on the momentum correlation that can be observed in the far field plane $\pi'$,
by considering two symmetrical detection areas. The variance of $N_-$ normalized to shot noise
is plotted in Fig.\ref{fig8} as a function of the detector size $d$ and for different values
of the pump beam waist. The result is similar to the one obtained for the type I crystal configuration
illustrated in Fig.\ref{fig7}: fluctuations are well below shot noise
only if $d$ is larger
than the characteristic resolution length of the system $x_{diff}=\frac{\lambda f}{2\pi}\delta q_0$.
We note that the number of temporal modes that are amplified by the
crystal is much lower than in the degenerate type I configuration, since in
this case temporal walk-off between the signal
and the idler fields reduces drastically the emission bandwidth $\Omega_0$.
Indeed, for the picosecond pump pulse we considered here
$\delta \omega_0/\Omega_0$ is in the order of the unity for the type II configuration, against the $~0.02$ value found
for the type I LBO crystal at degeneracy, for which temporal walk-off is not present.
The numerical simulations show that this feature of type II phase-matching does not affect
spatial correlations, but simply lowers the number of generated photon pairs per pulse.

\subsection{Near field correlation in type II crystals}
\label{numtypeIIn}
A main advantage of the type II configuration lies in the fact that
the signal and the idler fields have different
polarizations and can therefore be manipulated more easily. In
particular, it is possible to measure their mutual correlation in the
near field after they have been physically separated by a polarizing
beam splitter, as shown schematically in Fig.~\ref{fig4}.
The lenses $L$ and $L'$ shown in the figure simply perform
the $2f-2f$ imaging of the "near field plane" $\pi$ onto the two detection planes.
For the moment, we only assume that plane $\pi$ is located at some coordinate $z$ inside the crystal.
The S/I field self- and cross-correlation functions,
which in the plane-wave pump limit have the expressions (\ref{g11}) and (\ref{g12}),
display pronounced peaks for $\vec{x}~'=\vec{x}$. In particular,
the cross-correlation peak of $\Gamma_{12}(\vec{x},\vec{x}~',\Omega)$
describes the position entanglement of the S/I
photons, which are generated in pairs in the same region
of the crystal. The width of the peaks is on the order of
the coherence length $x_{coh}=1/q_0$ defined in Eq.(\ref{xcoh}).
It reflects the spread out of the generated photons due to diffraction,
which increases proportionally to the square root of the propagation distance.
In addition, depending on the phase-matching conditions, twin
photons can be emitted non collinearly with an aperture angle on the order
of $\alpha_R=2q_R/\overline{k}$, $q_R$ being the radius of the rings in Fourier space given by Eq.(\ref{qRqC}).
This introduce a further indeterminacy on the order of
$\alpha_R l_c/2=q_R/q_0 x_{coh}$
in the relative positions of the twin photons measured in the near field.
We expect therefore that the two macroscopic fields display identical fluctuations
when observed from the same region of the near field plane, provided that the detection areas
$R_1$ and $R_2$ are larger than the indeterminacy introduced by these propagation effects.
However, for crystal lengths on the order of a few millimeters
this indeterminacy can be as large as several tens to hundreds of microns, so that a substantial
portion of the two beams must be intercepted in order to measure significant correlation
in the quantum domain.
We shall see with a specific example that this difficulty can be at least partially
overcome.

The role of spatial walk-off is more subtle to determine.
As pointed out in \cite{koch}, the Poynting vectors of the phase-matched modes,
which determine the photon fluxes of the signal and the idler
beams, generate two cones which have the same axis inside the
crystal.
To verify this point explicitly we consider the Poynting vectors associated
to two particular modes  of the signal and idler fields at $\Omega=0$
with transverse wave-vectors $\vec{q}_1$ and $\vec{q}_2$.
Within the paraxial approximation their directions are determined by the
(two dimensional) angles
\begin{equation}
\vec{\alpha}_1=\frac{\vec{q}_1}{k_1}\;,~~~~~
\vec{\alpha}_2=\frac{\vec{q}_2}{k_2}+\vec{\rho}_2,
\end{equation}
where $\vec{\rho}_2\equiv(0,-\rho_2)$ indicates the walk-off direction of the idler field (the minus sign
is due to the fact that in the chosen reference frame the walk-off is oriented
opposite to the $y$-axis, $\rho_0$ and $\rho_2$ being assumed to be positive).
For the modes propagating along the axis of the signal and idler
cones, we have $\vec{q}_1=(0,-q_C)$ and
$\vec{q}_2=(0,q_C)$ with $q_C=\frac{1}{2}\overline{k}\rho_2$,
from which we see that the corresponding Poynting vectors are
collinear with $\vec{\alpha}_1=\vec{\alpha_2}=(0,-\frac{k_2}{k_1+k_2}\rho_2)$.
As a result, if the near field is measured directly
on the crystal output face $z=l_c$, walk-off
do not contribute to the indeterminacy in the
relative position of the twin photons. It should be stressed, however, that
free propagation beyond the crystal occurs at angles which are simply proportional to
the transverse wave-vector of the phase-matched modes,
since in free space the Poynting vector and the $k$-vector directions again coincide.
Outside the crystal the signal and idler emission cones are therefore oriented
along different directions with an aperture angle $\alpha_C=2q_R/k^{(v)}=\frac{\overline{k}}{k^{(v)}}\rho_2$,
as illustrated schematically in Fig.\ref{fig1}a, and give rise to separate rings in the far field
($k^{(v)}=2\pi/\lambda_1=2\pi/\lambda_2$ denote here the wave number of the S/I fields in free space
at the career frequencies).
In the simulation illustrated in Fig.\ref{fig9}, we consider the type II BBO
crystal in the same conditions described in Sec.\ref{numtypeIIf}.
The near field coherence length is $x_{coh}=16.6\mu$m and $q_R=0$.
The plot displays the noise reduction factor
$\langle (N_-)^2 \rangle_{\pi}/\langle N_+\rangle_{\pi}$,
evaluated numerically as a function of the 1-D detector size $d$.
If the near field observation plane $\pi$ coincides with the output face of
the crystal at $z=l_c$ (white circle), we see that the
fluctuations are significantly reduced only when $d$ is about 15 times
larger than $x_{coh}$. The improved result (black
squares) has been obtained by imaging onto the detection planes
a plane inside the crystal at $z=l_c-\Delta z$, rather than
the crystal output face. Furthermore, the array of pixel detectors in the signal and idler arms
are shifted with respect to each other by a distance $\Delta y$
in the transverse direction of
walk-off.  Notice that this is a 1-D simulation, and the quantities $\Delta y$ and $\Delta z$ correspond
to the quantities $\vec{x}_1 -\vec{x}_2 $ and $l_c-z$, respectively, which appears in the definition (\ref{H12}))
of the function $H_{12}(\vec{q},\vec{q}~')$.
$\Delta z$ and $\Delta y$ are chosen in order
to minimize the dependence on $\vec{q}$ and $\vec{q}\,'$ of the phase of the integrand in Eq.(\ref{cor2}),
maximizing in this way the photon-number cross correlation $\langle \delta N_1 \delta N_2 \rangle$.
This is achieved by taking
\begin{mathletters}
\label{dzdy}
\begin{eqnarray}
\Delta z_{opt} &=& \frac{\tanh\sigma_pl_c}{2\sigma_pl_c} \, \, \frac{n_1+n_2}{2n_1n_2} l_c\;,\label{dz}\\
\Delta y_{opt} &=& \frac{\tanh\sigma_pl_c}{2\sigma_pl_c} \, \, \rho_2 l_c\;.\label{dy}
\end{eqnarray}
\end{mathletters}
Indeed,
with these values of $\Delta y$ and $\Delta z$
the phase factor of the function $H_{12}(\vec{q},\vec{q}\,')$ defined by Eq.(\ref{H12})
nearly cancels the phase of the gain function product appearing in the r.h.s. of
Eq.(\ref{cor2}), as can be verified by using the approximate expression
\begin{mathletters}
\begin{eqnarray}
\label{argu1v2}
\arg\left[U_1(\vec{q},\Omega)V_2(-\vec{q},-\Omega)
          U_1^*(\vec{q}~',\Omega)V_2^*(-\vec{q}~',-\Omega)\right]
&\approx&
\frac{\tanh\sigma_pl_c}{2\sigma_pl_c}
\left[\Delta(\vec{q},\Omega)-\Delta(\vec{q}~',\Omega)\right]l_c\;,\\
&=&
-\frac{\tanh\sigma_pl_c}{2\sigma_pl_c}\left[\rho_2(\vec{q_y}-\vec{q_y}')
                                       +\frac{q^2-q'^2}{q_0}\right]
\end{eqnarray}
\end{mathletters}
which holds in the high gain region of the spatial frequency plane.
These shifts of the detection S/I planes are necessary in order to minimize both
the effects of diffraction and of spatial walk-off.

The plot of Fig.\ref{fig10} displays the variance of $N_-$ normalized to shot noise
in the $(\Delta z,\Delta y)$ plane, as calculated from Eqs.(\ref{cor}).
In this example, $d$ is only twice the coherence length, $x_{coh}=16.6\mu$m, while
$\Delta y_{opt}=47.5\mu$m and $\Delta z_{opt}=407\mu$m (with $\sigma_p l_c=3$).
The fluctuations are well above shot noise everywhere
except in the narrow diagonal region around the point $(\Delta z_{opt},\Delta y_{opt})$,
where $\langle (\delta N_-)^2\rangle_{\pi}/\langle N_+\rangle_{\pi}=0.3$.
Outside this region, the self-correlation become much larger than the cross-correlation and
the variance of $N_-$ rapidly exceeds the shot noise level.
This means that for such small detectors
a highly precise imaging of the near field determined by Eqs.(\ref{dzdy})
is therefore necessary in order to observe some noise reduction effect.

Although the previous results rely on a detailed description of the relative phases
of the signal and idler fields,
we can give them a more intuitive explanation based on the particle picture:

1) In the low gain regime ($\sigma_p l_c \ll 1$), photon pairs are generated uniformly along the crystal.
In this case the choice $\Delta z\approx l_c/2$ lowers the effect of diffraction and non collinear
propagation, since the mean propagation distance that
photon pairs must undergo to reach the imaging plane $\pi$ from the point
in which they are created is minimized.
The factor depending on the refractive indexes takes into account
that photons propagate in a dense medium rather than in free space.

2) On the other side, the fields imaged from a generic plane inside the crystal
lying at distance $\Delta z$ from the output face
can be obtained by a virtual free space back-propagation
from plane $z=l_c$ to plane $z=l_c-\Delta z$. As already mentioned, free space
propagation leads to an angular divergence of the signal and idler beams
with a mean aperture angle equal to $\alpha_C\approx\rho_2$ along the walk-off direction.
The signal and idler photons are therefore pull apart a distance
$\Delta y=\alpha_C\Delta z\approx \rho_2\Delta z$ when observed in plane $z=l_c-\Delta z$.
In order to compensate this effect the two
detection areas must be separated by the same distance along the the $y$-axis.
This explain why the value of $\Delta y$ for which the fluctuations are minimized
in a given imaging plane is proportional to $\Delta z$, as it appears from
the cigar shaped region of low fluctuations along the diagonal direction illustrated in
the contour plot of Fig.~\ref{fig10}. In particular for $\Delta z=\Delta z_{opt}$ walk-off is compensated by taking
$\Delta y=\alpha_C\Delta z_{opt}=\rho_2 l_c/2$ and we obtain thereby
the shifts given by Eqs.~(\ref{dz}) and (\ref{dy}) for the limit $\sigma_p l_c\ll 1$.

3) The factor depending on the gain parameter in Eq.(\ref{dz}), which decreases as $\sigma_p l_c$ increases,
can be understood by noting that in a high gain regime most of the
photon pairs are generated in the last part of the crystal,
because of a cascading effect. Hence in order to minimize the propagation distance from the point where
they are created to the plane $\pi$, this plane should be taken closer and closer
to the crystal exit face as the gain is increased.

Thank to this procedure, the noise in $N_-$ is considerably lowered with
respect to measurements performed at plane $z=l_c$ with aligned detection areas
(i.e. with $\Delta z=\Delta y=0$).
Although the values (\ref{dz}) and(\ref{dy}) have been evaluated within the plane-wave pump
approximation, the numerical simulations demonstrate that the procedure
works well even when the pump has a finite size.
It should also be emphasized that both shifts (\ref{dz}) and (\ref{dy})
proved to be equally necessary in order to obtain this improvement.

\section{Conclusions}
The results of this paper demonstrate that SPDC is able to display
spatial correlation effects at the level of
quantum fluctuations even in a regime of
high gain, i.e. when the down-converted photons form macroscopic fields.
The quantum origin of the S/I correlations, which lie in the position
and momentum entanglement of the twin photons
building up the two beams, can be best demonstrated
in a type II phase-matching configuration where both near field and far
field measurements can be implemented.
We showed numerically that spatial far field correlations of quantum
origin are observable when
the pump beam waist is in the millimeter range and the detection areas are larger
than the resolution area of the system.
Near field correlations seem more difficult to observe experimentally since
propagation tends to destroy the position entanglement of the generated photon pairs.
We proposed a detection scheme which allows to optimize their measurement by
compensating the detrimental effect of diffraction.
These results are strongly related to a recent paper of ours \cite{Qent},
which discusses the topic of entangled imaging and extends this technique
to the macroscopic domain. The simultaneous presence of spatial entanglement in both
the near and the far field plays a crucial role in the analysis of \cite{Qent}.
In this paper we provide a more quantitative analysis of the level of quantum correlation
which is present in the far field on the one hand and in the near field on the other.

\section*{Acknowledgments}
This work is supported by the European FET  Project QUANTIM
(Quantum Imaging). We are grateful to Paolo Di Trapani, Ottavia
Jedrkiewicz and Yunkun Jiang for their precious collaboration.
\appendix
\section{Input-output formalism}\label{appendixA}
The finite pump pulse bandwidth in space and in time
generates coupling between all modes of the S/I field, deteriorating
thereby the perfect correlation between the $(\vec{q},\Omega)_1$ and $(-\vec{q},-\Omega)_2$ modes.
In this section we generalize the input-output formalism of the PWPA
to the finite pump case.
Input-output transformations (\ref{inputoutput}) are replaced by the more general
linear transformation
\begin{mathletters}
\begin{eqnarray}
\label{inputoutput2}
a_1(z,\vec{q},\Omega)
&=&\int d\vec{q}~'\int d\Omega'
\left[
{\cal{U}}_1(z;\vec{q},\Omega;\vec{q}~'\Omega')a_1^{in}(\vec{q}+\vec{q}~',\Omega+\Omega')
+{\cal{V}}_1(z;\vec{q},\Omega;\vec{q}~',\Omega')a_2^{in\dag}(-\vec{q}+\vec{q}~',-\Omega+\Omega')
\right]\;,\\
a_2(z,\vec{q},\Omega)
&=&\int d\vec{q}~'\int d\Omega'
\left[
{\cal{V}}_2(z;\vec{q},\Omega;\vec{q}~',\Omega')a_1^{in\dag}(-\vec{q}+\vec{q}~',-\Omega+\Omega')
+{\cal{U}}_2(z;\vec{q},\Omega;\vec{q}~'\Omega')a_2^{in}(\vec{q}+\vec{q}~',\Omega+\Omega')
\right]\;,
\end{eqnarray}
\end{mathletters}
which express the fields in a generic plane $z$ inside the crystal in the form of a convolution integral
with the input field Fourier modes $a_{j}^{in}(\vec{q},\Omega)=a_j(z=0,\vec{q},\Omega)$, $j=1,2$.
>From Eqs.(\ref{waveq}), we can obtain a fully equivalent set
of equations for the propagation kernels
\begin{mathletters}
\label{kernel}
\begin{eqnarray}
\frac{\partial {\cal{U}}_1(z;\vec{q},\Omega;\vec{q}~',\Omega')}{\partial z}&=&
    i\delta_1(\vec{q},\Omega)
    {\cal{U}}_1(z;\vec{q},\Omega;\vec{q}~',\Omega')\label{kernela}\\
    & &+\sigma e^{-i\Delta_0 z}
    \int\frac{d\vec{q}~''}{2\pi} \int\frac{d\Omega''}{\sqrt{2\pi}}\;
    A_0(z,\vec{q}~'',\Omega''){\cal{V}}_2^*(z;\vec{q}~''-\vec{q},\Omega''-\Omega;\vec{q}~'+\vec{q}~'',\Omega'+\Omega'')\;,\nonumber\\
\frac{\partial {\cal{V}}_2(z;\vec{q},\Omega,\vec{q}~',\Omega')}{\partial z}&=&
    i\delta_2(\vec{q},\Omega)
    {\cal{V}}_2(z;\vec{q},\Omega,\vec{q}~',\Omega')\label{kernelb}\\
    & &+\sigma e^{-i\Delta_0 z}
    \int\frac{d\vec{q}~''}{2\pi} \int\frac{d\Omega''}{\sqrt{2\pi}}\;
    A_0(z,\vec{q}~'',\Omega''){\cal{U}}_1^*(z;\vec{q}~''-\vec{q},\Omega''-\Omega;\vec{q}~'+\vec{q}~'',\Omega'+\Omega'')\;.\nonumber
\end{eqnarray}
\end{mathletters}
The equations for the remaining kernels
${\cal{V}}_1$ and ${\cal{U}}_2$ can be obtained by interchanging indexes 1 and 2,
and the following initial conditions must be fulfilled
\begin{mathletters}
\label{init0}
\begin{eqnarray}
&&{\cal{U}}_{j}(z=0;\vec{q},\Omega;\vec{q}~',\Omega')=\delta(\vec{q}~')\delta(\Omega')\;,\\
&&{\cal{V}}_{j}(z=0;\vec{q},\Omega;\vec{q}~',\Omega')=0\;,
\hspace{2cm}(j=1,2)\;.
\end{eqnarray}
\end{mathletters}
It can be shown that the solutions of this set of
equations satisfy the relations
\begin{mathletters}
\label{unitarity2}
\begin{eqnarray}
\int d\vec{q}~'' \int d\Omega'' & &
\left[
   {\cal{U}}_1(z;\vec{q},\Omega;\vec{q}~''-\vec{q},\Omega''-\Omega)
   {\cal{U}}_1^*(z;\vec{q},\Omega;\vec{q}~''-\vec{q}~',\Omega''-\Omega')
\right.\\
& &\left.
  -{\cal{V}}_1(z;\vec{q},\Omega;\vec{q}~''+\vec{q},\Omega''+\Omega)
   {\cal{V}}_1^*(z;\vec{q},\Omega;\vec{q}~''+\vec{q}~',\Omega''+\Omega')
\right]
=\delta(\vec{q}~'-\vec{q})\delta(\Omega-\Omega')\;,\nonumber\\
\int d\vec{q}~'' \int d\Omega''& &
   {\cal{U}}_1(z;\vec{q},\Omega;\vec{q}~''-\vec{q},\Omega''-\Omega)
   {\cal{V}}_2(z;\vec{q},\Omega;\vec{q}~''+\vec{q}~',\Omega''+\Omega')\\
& &=\int d\vec{q}~'' \int d\Omega''
   {\cal{V}}_1(z;\vec{q},\Omega;\vec{q}~''+\vec{q},\Omega''+\Omega)
   {\cal{U}}_2(z;\vec{q},\Omega;\vec{q}~''-\vec{q}~',\Omega''-\Omega')\;,\nonumber
\end{eqnarray}
\end{mathletters}
which generalize the unitarity conditions (\ref{unitarity}) beyond the case of
plane-wave and cw pump.

The normally ordered photon number correlation function
in a generic transverse plane $z$ can be written as
\begin{mathletters}
\begin{eqnarray}
\label{corr}
G_{ij}^{(z)}(\vec{x},t,\vec{x}~',t')&=&
\langle a_i^{\dag}(z,\vec{x},t) a_j^{\dag}(z,\vec{x}~',t')a_j(z,\vec{x}~',t')a_i(z,\vec{x},t) \rangle
-\langle a_i^{\dag}(z,\vec{x},t)a_i(z,\vec{x},t)\rangle
\langle a_j^{\dag}(z,\vec{x}~',t')a_j(z,\vec{x}~',t')\rangle\label{corra}\\
&=&
|\langle a_i^{\dag}(z,\vec{x},t)a_j(z,\vec{x}~',t')\rangle|^2
+|\langle a_i(z,\vec{x},t)a_j(z,\vec{x}~',t')\rangle|^2\;.\label{corrb}
\end{eqnarray}
\end{mathletters}
In the last identity we made use of the general property
characterizing fields with a gaussian statistics,
which allows to write the fourth order field correlations as a sum of
products of the second order correlation functions (see e.g. \cite{gardiner}).
We now consider explicitly the transformations relating the S/I fields in the planes where
detection is performed, to those on the crystal output face, $a_j^{out}(\vec{q},t)$:
\begin{equation}
\label{fresnel2}
a_j(z,\vec{x},t)=\int d\vec{q}~h_j(\vec{x},\vec{q}) a_j^{out}(\vec{q},t)
                =\int d\vec{x}~'
                    ~h_j(\vec{x},\vec{x}~') a_j^{out}(\vec{x}~',t)~~~~~~~~(j=1,2)\;,
\end{equation}
with
\begin{equation}
\label{hxq}
h_j(\vec{x},\vec{q})\equiv\int \frac{d\vec{x}~'}{2\pi} e^{i\vec{q}\cdot\vec{x}~'}h_j(\vec{x},\vec{x}~')
\end{equation}
Propagation outside the crystal can include several optical devices, such as lenses and
polarizing beam splitters, and can take different paths for the signal and idler beams
(see scheme of Fig.\ref{fig4}).
We shall assume however that it occurs without losses and
this latter condition implies that the Fresnel kernels $h_j(\vec{x},\vec{q})$
satisfy the relation
\begin{equation}
\label{kernelxq}
\int d\vec{x} ~h_j^*(\vec{x},\vec{q})h_j(\vec{x},\vec{q}~')=\delta(\vec{q}-\vec{q}')\;,
~~~~~~(j=1,2)\;,
\end{equation}
which can be obtained by requiring that the commutation rules (\ref{commut}) are
preserved in the transformation.
When the detection time $T_d$ is much longer than the coherence time $\Omega_0^{-1}$,
the photon number self- and cross-correlations measured over two detection
areas $R_1$ and $R_2$, as defined by Eq.(\ref{dnij}), can be written as
\begin{mathletters}
\label{corrd}
\begin{eqnarray}
\langle :(\delta N_1)^2:\rangle_z&=&
\int d\Omega \int d\Omega '
\int_{R_1} d\vec{x}
\int_{R_1} d\vec{x}~'
\left|\int d\vec{q} \int d\vec{q}~'h_1^*(\vec{x},\vec{q})h_1(\vec{x}~',\vec{q}~')
              \langle a_1^{out\dag}(\vec{q},\Omega)a_1^{out}(\vec{q}~',\Omega')\rangle\right|^2\;,\\
\langle \delta N_1 \delta N_2 \rangle_z&=&
\int d\Omega \int d\Omega '
\int_{R_1} d\vec{x}\int_{R_2} d\vec{x}~'
\left|\int d\vec{q} \int d\vec{q}' h_1(\vec{x},\vec{q})h_2(\vec{x}~',\vec{q}~')
                          \langle a_1^{out}(\vec{q},\Omega)a_2^{out}(\vec{q}~',\Omega')\rangle\right|^2\;,
\end{eqnarray}
\end{mathletters}
where we used relation (\ref{fresnel2}) in order to express the second order field correlations
appearing in Eqs.(\ref{corrb}) in terms of the output field operators $a_j^{out}(\vec{q},\Omega)$, $j=1,2$.
The correlation functions of the output fields
can be expressed in terms of the
propagation kernels defined by Eq.(\ref{inputoutput2}) evaluated at plane $z=l_c$ as
\begin{mathletters}
\begin{eqnarray}\label{corr1}
\langle a_1^{out\dag}(\vec{q},\Omega) a_1^{out}(\vec{q}~',\Omega') \rangle
  &=&\int d\vec{x} \int dt ~e^{~i(\vec{q}-\vec{q}~')\cdot\vec{x}-i(\Omega-\Omega)t}
              ~{\cal{V}}_1^*(l_c;\vec{q},\Omega;\vec{x},t)
              {\cal{V}}_1(l_c;\vec{q}~',\Omega';\vec{x},t)\;,\label{corr1a}\\
\langle a_1^{out}(\vec{q},\Omega) a_2^{out}(\vec{q}~',\Omega') \rangle
  &=&\int d\vec{x} \int dt ~e^{~i(\vec{q}+\vec{q}~')\cdot\vec{x}-i(\Omega+\Omega)t}
              ~{\cal{U}}_1^*(l_c;\vec{q},\Omega;\vec{x},t)
              {\cal{V}}_2(l_c;\vec{q}~',\Omega';\vec{x},t)\;,\label{corr1b}\\
\langle a_1^{out}(\vec{q},\Omega) a_1^{out}(\vec{q}~',\Omega') \rangle&=&
\langle a_1^{out\dag}(\vec{q},\Omega) a_2^{out}(\vec{q}~',\Omega') \rangle=0\;,
\end{eqnarray}
\end{mathletters}
with
\begin{equation}
\label{fourieruv}
{\cal{U}}_{j}(z;\vec{q},\Omega;\vec{x},t)
=
\int \frac{d\vec{q}}{2\pi} \int \frac{dt}{\sqrt{2\pi}}
 e^{i\vec{q}~'\cdot \vec{x}-i\Omega' t}
{\cal{U}}_{j}(z;\vec{q},\Omega;\vec{q}~',\Omega')~~~~~~~(j=1,2)\;.
\end{equation}
A similar definition holds for the functions ${\cal{V}}_{j}(z;\vec{q},\Omega;\vec{x},t)$, $j=1,2$.
In case $R_1$ and $R_2$ intercept all the photons of the signal and the idler fields generated
in the down-conversion process, it can be shown that the variance of $N_-=N_1-N_2$ vanishes
if condition (\ref{kernelxq}) is fulfilled (i.e if free propagation occurs without losses).
However, in general such a result does not
hold if the two detectors collect photons only from finite portions of the two beams.

\section{Approximate solution in the quasi-stationary regime}\label{appendixB}
In this appendix we derive an approximate analytical solution of the propagation equations
(\ref{waveq}) assuming the spatial and temporal frequency
bandwidths of the pump are small but finite; more precisely we assume that
conditions (\ref{largepump}) are satisfied.

It is useful to write the propagation equations in the reference
frame comoving with the pump field envelope, whose coordinates are related
to the original laboratory coordinates through the
linear transformation $t'=t-k_0'z$, $y'=y+\rho_0 z$.
In Fourier space this correspond to multiplying the Fourier components of the S/I field envelopes
by $e^{-i\left[k_0'\Omega+\rho_0q _y\right]z}$. More precisely we consider the transformation
\begin{equation}
a_j'(z,\vec{q},\Omega)=e^{i\left[\frac{\Delta_0}{2}-k_0'\Omega-\rho_0q _y\right]z}a_j(z,\vec{q},\Omega)\;,~~~~~~~(j=1,2)\;,
\end{equation}
where the constant phase factor $e^{i\frac{\Delta_0}{2}z}$ has been added
in order to eliminate $e^{i\Delta_0 z}$ from the convolution term in Eqs.(\ref{kernel}).
Input-output transformations (\ref{inputoutput2}) and  unitarity conditions (\ref{unitarity2})
still hold for the transformed kernels
\begin{mathletters}
\begin{eqnarray}
&&{\cal{U}}_j\,'(z;\vec{q},\Omega;\vec{q}~',\Omega')
 =e^{i\left[\frac{\Delta_0}{2}-k_0'\Omega-\rho_0q_y\right]z}
                {\cal{U}}_j(z;\vec{q},\Omega;\vec{q}~',\Omega')\;,\\
&&{\cal{V}}_j\,'(z;\vec{q},\Omega;\vec{q}~',\Omega')
 =e^{i\left[\frac{\Delta_0}{2}-k_0'\Omega-\rho_0q_y\right]z}
                {\cal{V}}_j(z;\vec{q},\Omega;\vec{q}~',\Omega')\;,~~~~(j=1,2),
\end{eqnarray}
\end{mathletters}
which satisfy the propagation equations
\begin{mathletters}
\label{kernel2}
\begin{eqnarray}
\frac{\partial {\cal{U}}_1\,'(z;\vec{q},\Omega;\vec{q}~',\Omega')}{\partial z}&=&
    i\delta_1'(\vec{q},\Omega)
    {\cal{U}}_1\,'(z;\vec{q},\Omega;\vec{q}\,',\Omega')\label{kernel2a}\\
    & &+
    \int\frac{d\vec{q}~''}{2\pi} \int\frac{d\Omega''}{\sqrt{2\pi}}\;
    e^{i\delta_0'(\vec{q}\,'',\Omega'')z}
    A_0(z=0,\vec{q}~'',\Omega''){\cal{V}}_2^{'*}(z;\vec{q}~''-\vec{q},\Omega''-\Omega;\vec{q}~'+\vec{q}~'',\Omega'+\Omega'')\;,\nonumber\\
\frac{\partial {\cal{V}}_2\,'(z;\vec{q},\Omega,\vec{q}~',\Omega')}{\partial z}&=&
    i\delta_2'(\vec{q},\Omega)
    {\cal{V}}_2\,'(z;\vec{q},\Omega,\vec{q}~',\Omega')\label{kernel2b}\\
    & &+
    \int\frac{d\vec{q}~''}{2\pi} \int\frac{d\Omega''}{\sqrt{2\pi}}\;
    e^{i\delta_0'(\vec{q}\,'',\Omega'')z}
    A_0(z=0,\vec{q}~'',\Omega''){\cal{U}}_1^{\,'*}(z;\vec{q}~''-\vec{q},\Omega''-\Omega;\vec{q}~'+\vec{q}~'',\Omega'+\Omega'')\;.\nonumber
\end{eqnarray}
\end{mathletters}
The explicit form for the propagation of the pump field (\ref{prop0}) has been used
and we defined the new detuning parameters
\begin{mathletters}
\begin{eqnarray}
&&\delta_j'(\vec{q},\Omega)=\frac{\Delta_0}{2}+\delta_j(\vec{q},\Omega)-\rho_0 q_y -k_0'\Omega~~~~~~(j=1,2)\;,\\
&&\delta_0'(\vec{q},\Omega)=\frac{1}{2}k_0''\Omega^2-\frac{1}{2k_0}q^2\;.
\end{eqnarray}
\end{mathletters}
Remembering the hypothesis that the pump envelope Fourier
transform at plane $z=0$ has the gaussian form (\ref{pump0}) with $\delta q_0 \ll q_0$
and $\delta \omega_0 \ll \Omega_0$,
we can now apply the following approximations:

1) The phase term in the convolution
integrals can be neglected, since $\delta_0'(\vec{q},\Omega)l_c$ is at most on the order
of $\max\left\{\delta q_0^2/q_0^2,\delta\omega_0^2/\Omega_0^2\right\}$ in the region where
the Fourier transform of the pump envelope is not negligible.

2) We expect that the propagation kernels defined by Eq.(\ref{inputoutput2})
are characterized by the slow
variation scale $(q_0,\Omega_0)$ in their unprimed arguments, while
they are strongly peaked in the origin of the primed variable space $(\vec{q}~',\Omega')$,
in which they have a much faster variation scale.
This assumption is justified by the form of the solutions of Eqs.(\ref{kernel2})
obtained in the PWPA limit.
As $\delta q_0/q_0\rightarrow 0$ and $\delta\omega_0/\Omega_0\rightarrow 0$,
the primed propagation kernels satisfying initial conditions (\ref{init0}) take indeed the simple form:
\begin{mathletters}
\label{pwpasol}
\begin{eqnarray}
{\cal{U}}'_j(z;\vec{q},\Omega;\vec{q}~',\Omega')&=&\delta(\vec{q}~')\delta(\Omega')
                      e^{i\left[\frac{\Delta_0}{2}-k_0'\Omega-\rho_0q_y\right]z}U_j(z;\vec{q},\Omega)\;,\\
{\cal{V}}'_j(z;\vec{q},\Omega;\vec{q}~',\Omega')&=&\delta(\vec{q}~')\delta(\Omega')
                      e^{i\left[\frac{\Delta_0}{2}-k_0'\Omega-\rho_0q_y\right]z}V_j(z;\vec{q},\Omega)\;
\end{eqnarray}
\end{mathletters}
where the gain function $U_j(z;\vec{q},\Omega)$ and $V_j(z;\vec{q},\Omega)$
are given by Eqs.(\ref{uv}), with $l_c$ being replaced by the $z$-coordinate.
This latter hypothesis allows us to
neglect the dependence on $\vec{q}~''$ and $\Omega''$ in the first argument of the
kernels in the convolution integrals at the r.h.s. of Eqs.(\ref{kernel2a}) and(\ref{kernel2b}).
With these approximations, we obtain a system which can be solved analytically and which acquires
its simplest form when written for the kernels Fourier transformed in their primed
arguments (see definition (\ref{fourieruv}):
\begin{mathletters}
\label{waveq3}
\begin{eqnarray}
\frac{\partial {\cal{U}}_1\,'(\vec{q},\Omega;\vec{x},t)}{\partial z}&=&
    i\delta_1'(\vec{q},\Omega)
    {\cal{U}}_1\,'(\vec{q},\Omega;\vec{x},t)\\
    & &+\sigma
    A_0(z=0,\vec{x},t){\cal{V}}_2^{\,'*}(-\vec{q},-\Omega;-\vec{x},-t)\;,\nonumber\\
\frac{\partial {\cal{V}}_2\,'(\vec{q},\Omega,\vec{x},t)}{\partial z}&=&
    i\delta_2'(\vec{q},\Omega)
    {\cal{V}}_2\,'(\vec{q},\Omega,\vec{x},t)\\
    & &+\sigma
    A_0(z=0,\vec{x},t){\cal{U}}_1^{\,'*}(-\vec{q},-\Omega;-\vec{x},-t)\nonumber\;,
\end{eqnarray}
\end{mathletters}
The solution of this system satisfying initial conditions (\ref{init0}), which now read
${\cal{U}}~'_{j}(z=0;\vec{q},\Omega;\vec{x},t)=1/(2\pi)^{3/2}$,
${\cal{V}}~'_{j}(z=0;\vec{q},\Omega;\vec{x},t)=0$, $(j=1,2)$,
are
\begin{eqnarray}
\label{UVxt}
     {\cal{U}}_1\,'(z;\vec{q},\Omega;\vec{x},t)&=&
     \exp\left[i\frac{\delta_1'(\vec{q},\Omega)-\delta_2'(-\vec{q},-\Omega)}{2}z\right]
     U(z;\vec{q},\Omega;\vec{x},t)\;,\\
     {\cal{V}}_1\,'(z;\vec{q},\Omega;\vec{x},t)&=&
     \exp\left[i\frac{\delta_1'(\vec{q},\Omega)-\delta_2'(-\vec{q},-\Omega)}{2}z\right]
     V(z;\vec{q},\Omega;\vec{x},t)\;,\nonumber\\
     {\cal{U}}_2\,'(z;\vec{q},\Omega;\vec{x},t)&=&
     \exp\left[i\frac{\delta_2'(\vec{q},\Omega)-\delta_1'(-\vec{q},-\Omega)}{2}z\right]
     U(z;-\vec{q},-\Omega;-\vec{x},-t)\nonumber\;,\\
          {\cal{V}}_2\,'(z;\vec{q},\Omega;\vec{x},t)&=&
     \exp\left[i\frac{\delta_2'(\vec{q},\Omega)-\delta_1'(-\vec{q},-\Omega)}{2}z\right]
     V(z;-\vec{q},-\Omega;-\vec{x},-t)\;,\nonumber
\end{eqnarray}
with
\begin{eqnarray}
\label{uvxy}
& &U(z;\vec{q},\Omega;\vec{x},t)
  =\frac{1}{(2\pi)^{3/2}}
  \left[
     \cosh\Gamma(\vec{q},\Omega,\vec{x},t)z
     +i\frac{\Delta(\vec{q},\Omega)}{2\Gamma(\vec{q},\Omega,\vec{x},t)}
     \sinh\Gamma(\vec{q},\Omega,\vec{x},t)z
     \right]\;,\\
& &V(z;\vec{q},\Omega;\vec{x},t)
  =\frac{1}{(2\pi)^{3/2}}
     \frac{\sigma A_0(\vec{x},t)}{\Gamma(\vec{q},\Omega,\vec{x},t)}
     \sinh\Gamma(\vec{q},\Omega,\vec{x},t)z\;,\nonumber\\
& &\Gamma(\vec{q},\Omega;,\vec{x},t)
      =\sqrt{\sigma^2 A_0^2(\vec{x},t)-\frac{\Delta(\vec{q},\Omega)^2}{4}}\;,\\
& &\Delta(\vec{q},\Omega)=\delta_1'(\vec{q},\Omega)+\delta_2'(-\vec{q},-\Omega)
      =\Delta_0+\delta_1(\vec{q},\Omega)+\delta_2(-\vec{q},-\Omega)\;.
\end{eqnarray}
Clearly, as $\delta q_0,\delta \omega_0 \rightarrow 0$ these functions
function looselose their dependence on the space-time
coordinates $(\vec{x},t)$ and we recover the PWPA
solution expressed by Eqs.(\ref{uv}) and (\ref{Gamma}).
In the more general case in which the ratio $\delta q_0/q_0$ and $\delta q_0/q_0$ are small but finite,
the self-correlation function (\ref{corr1a}) is peaked at $\vec{q}'=\vec{q}$,
while the cross-correlation function (\ref{corr1b}) is peaked at $\vec{q}'=-\vec{q}$, the width
of both peaks being on the order of $\delta q_0$.
This is more clearly seen by considering the special limit in which $l_c\rightarrow 0$
and the gain parameter $\sigma A_p l_c$ remain a finite quantity.
In this case the effects of linear propagation, such as diffraction, dispersion
and walk-off become negligible and the propagation kernels (\ref{uvxy})
lose their dependence on $\vec{q}$ and $\Omega$ (as it can be inferred
from the fact that the characteristic bandwidths defined in Eq.(\ref{bandwidth})
go to infinity as $l_c\rightarrow 0$).
The correlation functions (\ref{corr1a}) and (\ref{corr1b}) are then simply the Fourier transforms
of sine and cosine hyperbolic function of $\sigma A_0(\vec{x},t) l_c$,
calculated in $\vec{q}-\vec{q}'$ and in $\vec{q}+\vec{q}'$ respectively.
We also verified that
that these approximate solutions looks very similar in shape
to those obtained with the complete numerical model
(see Fig.\ref{fig6}), although for the chosen crystal length of 4 mm
propagation effects are far from being negligible.

Considering the specific case of a far field measurement with the $f-f$ lens system
described in Sec.\ref{correlations},
the Fresnel kernels defined by Eq.(\ref{fresnel2}) take the form
$h_1(\vec{x},\vec{q})=h_2(\vec{x},\vec{q})=-\frac{2\pi i}{\lambda f}\delta
\left(\vec{q}-\frac{2\pi}{\lambda f}\vec{x}\right)$,
as it can be verified by substituting expression (\ref{farfieldb}) into Eq.(\ref{hxq}). Eqs.(\ref{corrd})
reduce then to
\begin{mathletters}
\begin{eqnarray}
&&\langle :(\delta N_1)^2:\rangle_{\pi'}=\langle :(\delta N_2)^2:\rangle_{\pi'}=
\int d\Omega \int d\Omega '
\int_{Q_1} d\vec{q}
\int_{Q_1} d\vec{q}~'
\left|\int d\vec{q} \int d\vec{q}~
              \langle a_1^{out\dag}(\vec{q},\Omega)a_1^{out}(\vec{q}~',\Omega')\rangle\right|^2\;,\\
&&\langle \delta N_1 \delta N_2 \rangle_{\pi'}=
\int d\Omega \int d\Omega '
\int_{Q_1} d\vec{q}\int_{Q_2}d\vec{q}~'
\left|\int d\vec{q} \int d\vec{q}'
                          \langle a_1^{out}(\vec{q},\Omega)a_2^{out}(\vec{q}~',\Omega')\rangle\right|^2\;,
\end{eqnarray}
\end{mathletters}
where $Q_1$ and $Q_2$ indicate the regions in the spatial frequency plane
corresponding to the two symmetrical detection areas $R_1$ and $R_2$ according to the mapping
$\vec{x}\rightarrow \frac{2\pi}{\lambda f}\vec{x}$. Knowing that $\langle (\delta N_-)^2\rangle
\rightarrow 0$ as $R_1,~R_2\rightarrow \infty$,
the localization of the cross- and self- correlation functions (\ref{corr1a}) and (\ref{corr1b})
on a area on the order of $\delta q_0^2$, for $\vec{q}'=\vec{q}$ and $\vec{q}'=-\vec{q}$
respectively, guarantees that nearly complete noise reduction is achieved if the area
of the two detectors is large compared to the resolution area determined by the pump
beam waist, that is $S_{diff}=(\lambda f/2\pi)^2\delta q_0^2$.

\noindent \vspace{1.cm}
\newpage
\begin{figure}

\epsfxsize=12cm
\centerline{\epsfbox{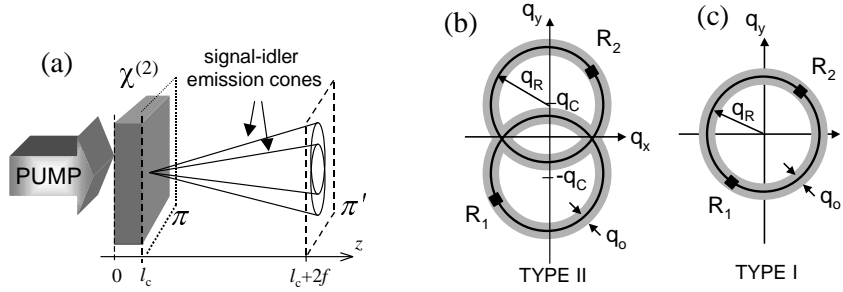}} \caption{Scheme
for the observation of spontaneous down-conversion in the far
field zone (a). The lens (not shown in the figure) is located at
$z=l_c+f$. (b) and (c) display the phase-matching curves
(\ref{rings}) in the spatial frequency plane for a type II (b) and
a type I (c) crystal respectively. The symmetrical black square
$R_1$ and $R_2$ indicate the location of the detectors from which
maximal signal-idler correlation can be measured.} \label{fig1}
\end{figure}
\noindent
\vspace{.5cm}
\begin{figure}[b]
\epsfysize=4.cm
\centerline{\epsffile{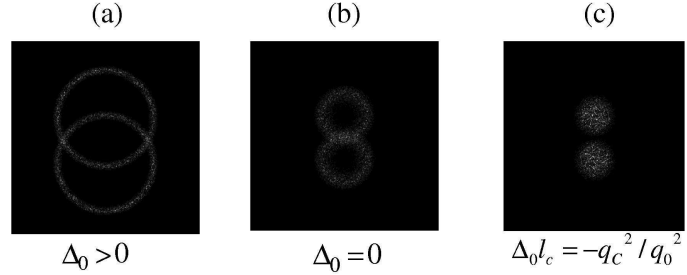}} \vspace{.2cm}
\caption{Typical far field pattern from spontaneous down
conversion in a type II crystal, assuming observation is performed
at the degenerate frequency (i.e. at $\lambda_1=\lambda_2$). They
are obtained for decreasing values of the collinear phase mismatch
parameter $\Delta_0$, which makes the radius of the rings shrinks
to zero.} \label{fig2}
\end{figure}
\vspace{.2cm}
\begin{figure}[t]
\epsfxsize=6cm \epsfysize=3.5cm
\centerline{\epsffile{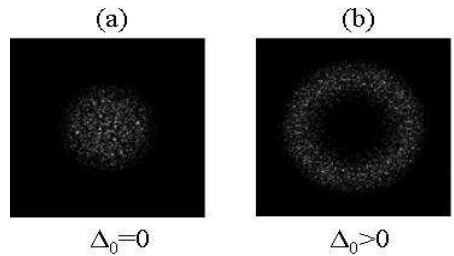}}
\caption{Far field pattern in a type I crystal at the degenerate
frequency for collinear (a) and non-collinear (b) phase-matching.}
\label{fig3}
\end{figure}
\begin{figure}
\vspace{1cm} \epsfxsize=8cm
\centerline{\epsfbox{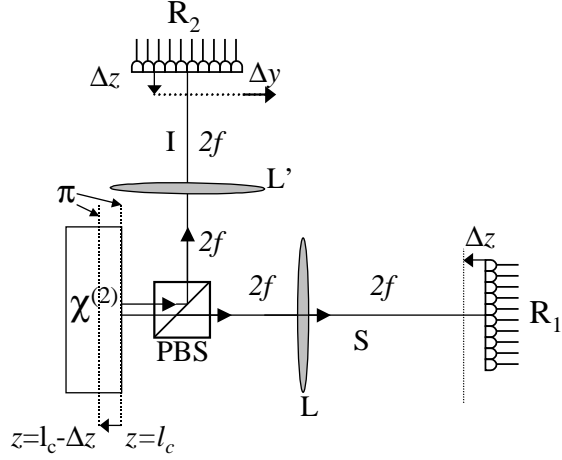}}
\caption{Detection scheme to measure spatial correlations in the
near field. A polarizing beam splitter (PBS) separates the S/I
beams. Their near fields, at the plane $\pi:z=l_c-\Delta z$, are
imaged by two lenses (L and L') onto the pixel detectors $R_1$ and
$R_2$, which lie in the plane conjugate to plane $\pi$. $\Delta z$
and $\Delta y$ indicate the spatial shifts applied to the optical
devices that are necessary to to optimize the measurement.}
\label{fig4}
\end{figure}
\begin{figure}[b]
\epsfysize=9cm
\centerline{\epsfbox{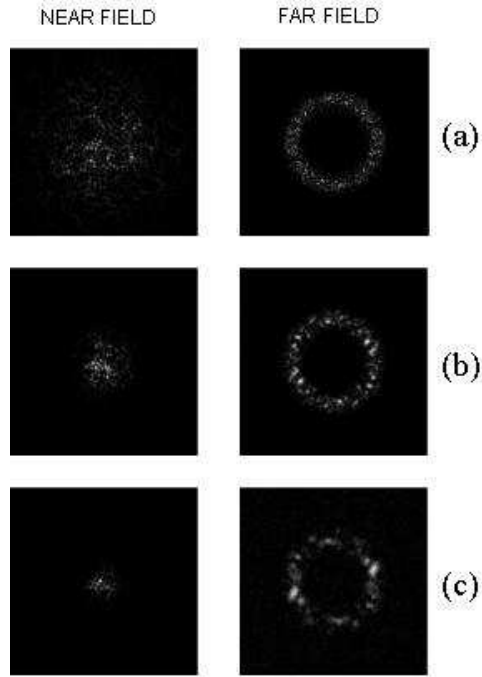}} \vspace{.5cm}
\caption{Near field (left) and far field (right) patterns obtained
for $\delta q_0/q_0=0.05$ (corresponding to a pump beam waist
$w_0=920\mu$m in the LBO case) (a), $\delta q_0/q_0=0.1$
($w_0=460\mu$m) (b), and $\delta q_0/q_0=0.3$ ($w_0=150\mu$m) (c).
$\Delta_0 l_c=13.6$ and $\sigma_p l_c=3.0$.} \label{fig5}
\end{figure}
\begin{figure}[b]
\vspace{1.7cm} \epsfxsize=11cm \centerline{\epsfbox{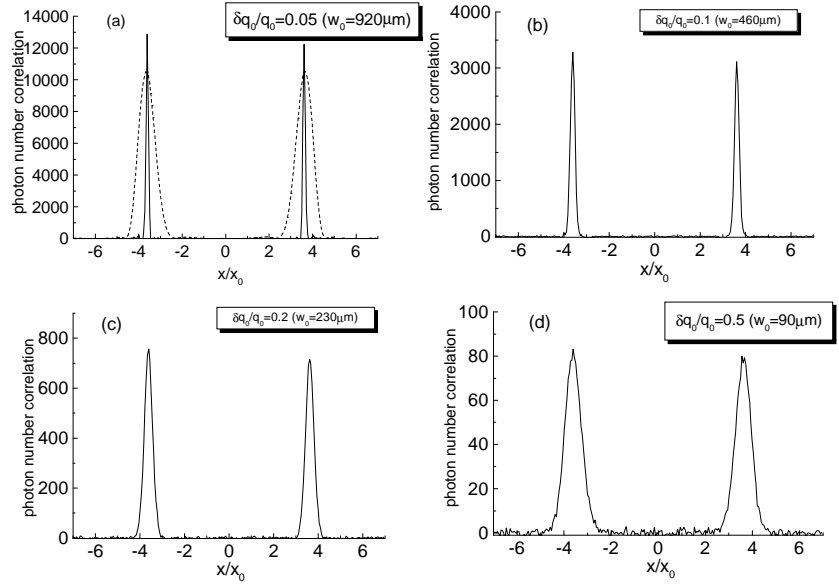}}
\caption{Far field correlations: $\langle :\delta N(\vec{x})\delta
N(\vec{x}~'):\rangle_{\pi'}$ is plotted as a function of $x$, for
increasing values of the ratio $\delta q_0/q_0$. In (a) the mean
photon number distribution profile is also shown (dotted line).
$x'=3.6\,x_0$ is kept fixed in the region of maximum gain. The
other parameters are the same as in Fig.\ref{fig5}.} \label{fig6}
\end{figure}
\begin{figure}[t]
\vspace{.3cm} \epsfxsize=9cm \centerline{\epsfbox{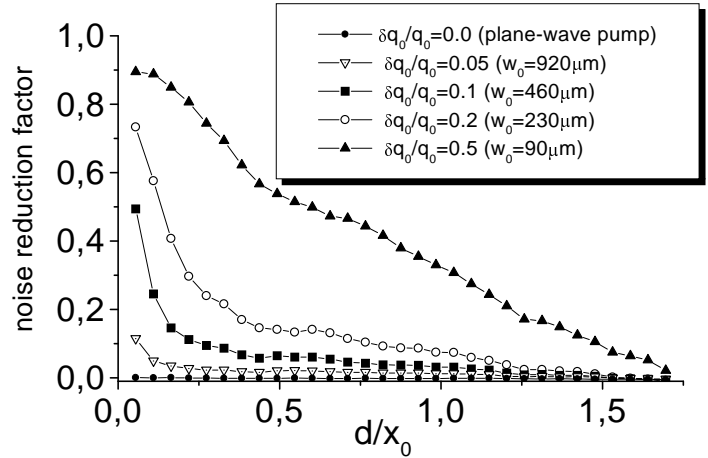}}
\caption{Far field correlation in type I: $\langle (\delta
N_-)^2\rangle_{\pi'}/\langle N_+ \rangle_{\pi'}$ is plotted as a
function of the detector size for different values of $\delta
q_0/q_0$. The other parameters are the same as in the previous
figure.} \label{fig7}
\end{figure}
\begin{figure}[t]
\epsfxsize=9cm
\centerline{\epsfbox{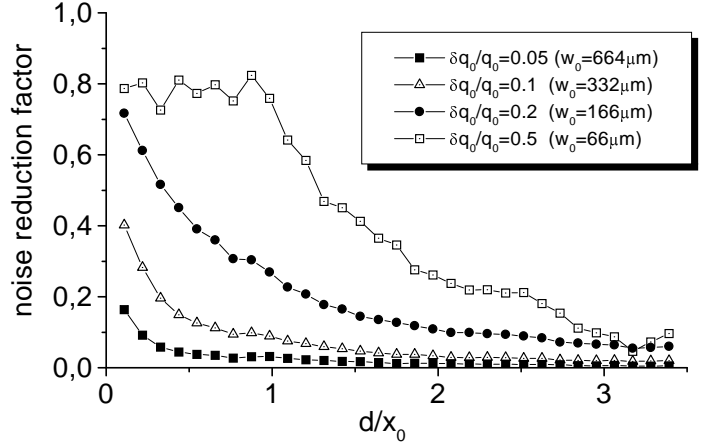}}
\caption{Far field
correlation: the ratio $\langle (\delta
N_-)^2\rangle_{\pi'}/\langle N_+\rangle_{\pi'}$ is plotted as a
function of the detector size $d$ for increasing value of the
ratio $\delta q_0/q_0$. The parametric gain is $\sigma_p l_c=4$.
The negative value of the collinear phase mismatch, $\Delta_0
l_c=-q_C^2/q_0^2=-74.4$, is such that the radii of the rings $q_R$
vanishes.} \label{fig8}
\end{figure}
\vspace{1.cm}
\begin{figure}[t]
\epsfxsize=9cm
\centerline{\epsfbox{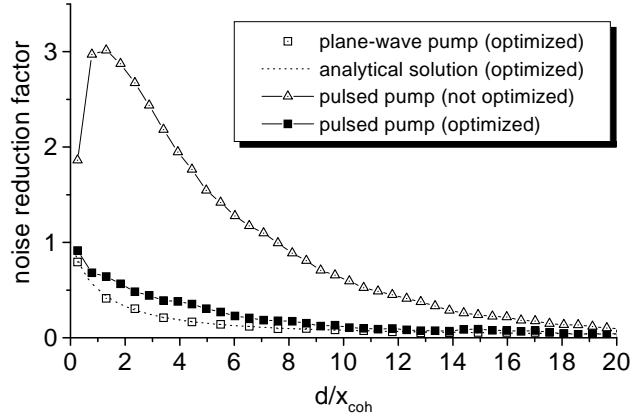}}   
\caption{Near field correlation: the ratio
$\langle (\delta N_-)^2\rangle_{\pi}/\langle N_+\rangle_{\pi}$ is plotted as a function of the
detector size. The parameters of the pulsed gaussian pump are
$w_0=332\mu$m ($\delta q_0/q_0=0.1$) and $\tau_0=1.5$~ps ($\delta
\omega_0/\Omega_0=1.14$); the gain is $\sigma_p l_c=3$ and $q_R=0$. The
simulations performed applying diffraction and walk-off
compensation (squares) are well below the one performed without
optimization (white triangle). The dashed line corresponds to the analytical solution
obtained in the PWPA, given by Eqs.(\ref{variance}) and (\ref{cor})}
\label{fig9}
\end{figure}
\begin{figure}[t]
\vspace{.3cm} \epsfxsize=9cm
\centerline{\epsfbox{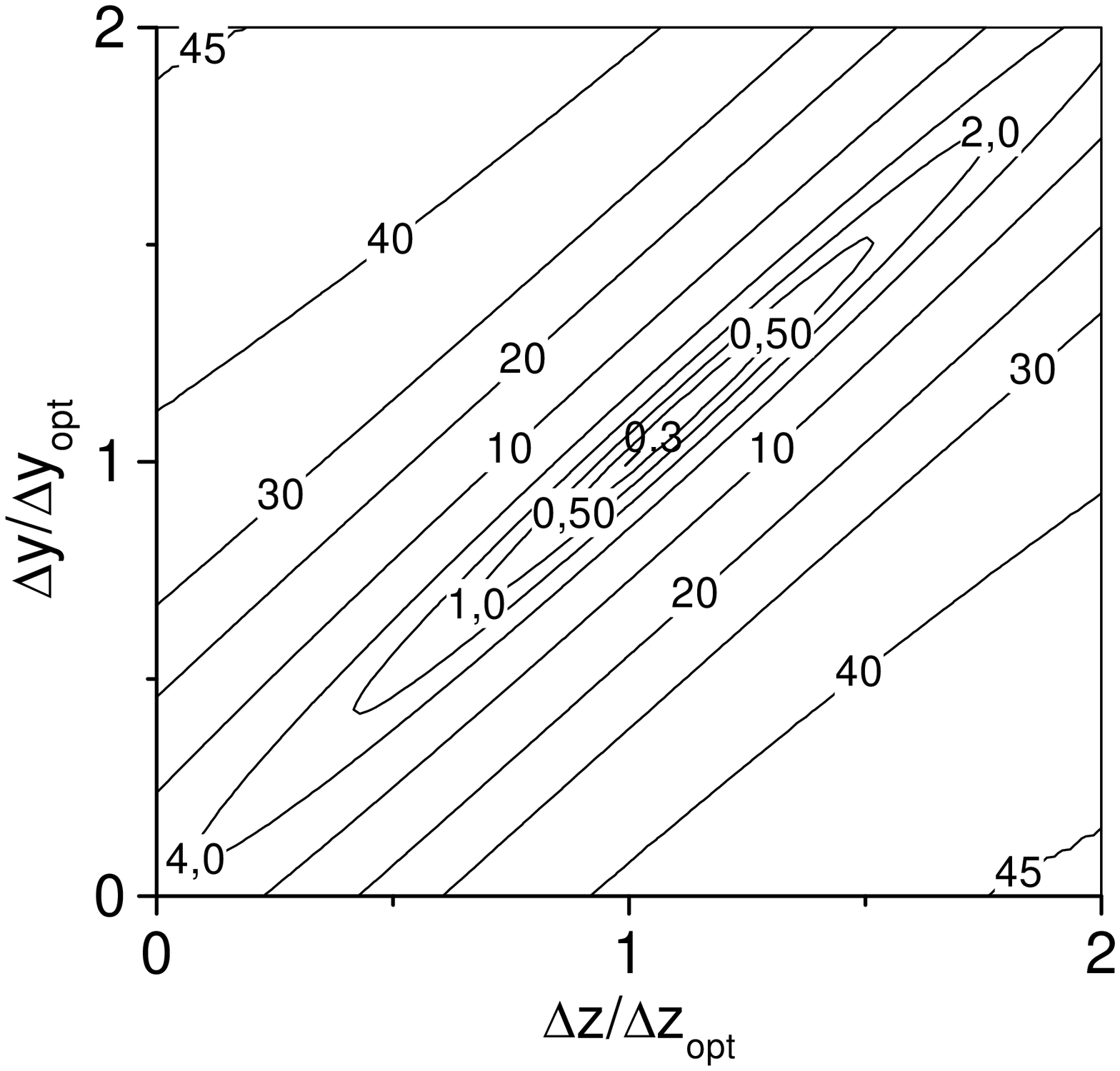}}
\caption{Near field correlation: contour plot of the ratio
$\langle (\delta N_-)^2\rangle_{\pi}/\langle N_+\rangle_{\pi}$ in
the $(\Delta z, \Delta y)$ plane calculated according to the PWPA
theory. The minimum in the center corresponds to the values given
by Eqs.(\ref{dzdy}).} \label{fig10}
\end{figure}


\begin{references}
\bibitem{advances} L.~A.~Lugiato, M.~Brambilla and A.~Gatti, in Advances in
                 Atomic, Molecular and Optical Physics, Vol.~40, p.~229,
                 Academic Press, Boston 1999.
\bibitem{review} M.~I.~Kolobov,{\it The spatial behavior of nonclassical light},
                 Rev.~Mod.~Phys. {\bf 71}, 1539 (1999) and references quoted therein.
\bibitem{aspects}A.~Gatti, E.~Brambilla, M.~I.~Kolobov and
                 L.~A.~Lugiato, J.~Opt.~B: Quant.~Semiclass.~Opt.
                 {\bf 2}, 196 (2000).
\bibitem{holo}   A.F. Abouraddy, B.E.A. Saleh, A.V. Sergienko, and M.C. Teich,
                 Optics Exp. {\bf 9}, 498 (2001);
\bibitem{telep}  I.V. Sokolov, M.I. Kolobov, A. Gatti and Lugiato,
                 Opt. Comm. {\bf 193},175 (2001)
\bibitem{fabre}  N. Treps, U. Andersen, B. Buchler, P. K. Lam, A. Maitre, H.-A. Bachor, and C. Fabre
                 Phys. Rev. Lett. {\bf 88}, 203601 (2002).
\bibitem{qimages}L.~A.~Lugiato, A. Gatti and E. Brambilla,
                 J. Opt. B: Quant. Semiclass. Opt. {\bf 4}, S176 (2002).
\bibitem{prl}   A.~Gatti, E.~Brambilla, L.~A.~Lugiato
                and M.~I.~Kolobov Phys.~Rev.~Lett.~ {\bf 83}, 1763 (1999).
\bibitem{fluor} E.Brambilla, A. Gatti, M. Kolobov and L.A.~Lugiato,
                Eur.~Phys.~J.~D {\bf 15},~117 (2001).
\bibitem{CCD}  Y.~Jiang, O.~Jedrkievicz, S.~Minardi, P.Di~Teapani, A.~Mosset, E.~Lantz
                 and F.Devaux, Eur.~Phys.~J.~D. {\bf 22} 521 (2003).
\bibitem{lantznum} E.~Lantz and F.~Devaux, Eur.~Phys.~J.~D. {\bf 17} 93 (2001).
\bibitem{kolobov89}  M.~I.~Kolobov and I.~V.~Sokolov, Sov.~Phys.~JETP
                 {\bf 69},1097 (1989); Phys.~Lett.~A {\bf 140},101 (1989).
\bibitem{handbook}  V.~G.~Dmitriev, G.~,G.~Gurzadyan, D.~N.~ Nikogosyan,
                {\em Handbook of nonlinear optical crystals}, Springer series in optical
                sciences, Springer-Verlag, Berlin (1991);
                N.~Boeuf et al, Optical Engineering, {\bf 39},~1016 (2000).
\bibitem{navez} P.Navez, E.Brambilla, A.Gatti and L.A.Lugiato,
                Phys.~Rev.~A {\bf 65},~13813 (2002).
\bibitem{lantz2} F.~Devaux and E.~Lantz,
                Eur.~Phys.~J.~D {\bf 8},~117 (2000).
\bibitem{ditrapani} A.~Berzanskis, W.~Chinaglia, L.~A.~Lugiato, K.~H.~Feller
                    and P.~Di Trapani, Phys.~Rev.~A {\bf 60}, 1626 (1999).
\bibitem{jost} B.~M.~Jost, A.~V.~Sergienko, A.~F.~Abouraddy, B.~E.~A.~Saleh
               and M.~C.~Teich, Opt. Express {\bf 3}, 81 (1998).
\bibitem{numrec} see e.g. W.~Press, B.~Flannery, S.~Teukolsky, W.~Vetterling
                 {\it Numerical Recipes}, Cambridge University Press (1992).
\bibitem{random} In the simulations we made use of a reliable gaussian random
number generator which is discussed in
R. Toral, A. Chakrabarti. Compute Physics Communications, 74 (1993) 327-334.
\bibitem{imaging} G.~Le Tolguenec, F.~Devaux and E.~Lantz,
                  Opt.~Lett.~{\bf 24},~1047 (1999);
                  F.~Devaux and E.~Lantz,
                  J.~Opt.~Soc.~Am.~B {\bf 12},~2245 (1995).
\bibitem{drummond} M.J.~Werner, M.G.~Raymer, M.~Beck and P.D.~Drummond
                   Phys.Rev.~A {\bf 52},~4202 (1995);
                   M.J.~Werner and P.D.~Drummond
                   Phys.Rev.~A {\bf 56},~1508 (1997).
\bibitem{gardiner} C. Gardiner, {\em Quantum noise}, Springer, Berlin (1991).
\bibitem{Zeilinger} A.~Zeilinger, Rev.~Mod.~Phys.{\bf 71}, S288 (1999).
\bibitem{koch} K.~Koch, E.C.~Cheung, G.T.~Moore,S.H.~Chakmakjian
and J.M.~Liu J.Of Quant.Electr. {\bf31},~769 (1995).

\bibitem{ditrapani3} A.~Gatti, L.~A.~Lugiato, G-L.~Oppo, R.~Martin,
                    P.~Di Trapani,and A.~Berzanskis, Opt. Expr.~{\bf 1}, 21 (1997). 
\bibitem{ditrapani2} P.~Di Trapani, A.~Andreoni, G.P.~Banfi, C.~Solcia, R.~Danelius
                     P.~Foggi, M.~Monguzzib and C.~Sozzi,
                     Phys.~Rev.~A {\bf 51   },~3164 (1995).
\bibitem{Qent} A.~Gatti, E.~Brambilla and L.~A.~Lugiato, Phys.~Rev.~Lett. {\bf 90}, 133603 (2003).
\end{references}
\end{document}